\documentclass[journal]{IEEEtran}

\usepackage[colorlinks=false,linkcolor=red,urlcolor=blue,
citecolor=gray,pdfborder={0 0 0}]{hyperref}

\usepackage[utf8]{inputenc}
\usepackage[dvipsnames]{xcolor}
\usepackage{graphicx}
\usepackage{amssymb,amsmath,amsfonts}
\usepackage{multicol,multirow}
\usepackage{booktabs}
\usepackage[printonlyused]{acronym}
\usepackage{xcolor}
\usepackage[capitalize]{cleveref}
\crefformat{equation}{(#2#1#3)}
\crefrangeformat{equation}{(#3#1#4)-(#5#2#6)}
\crefmultiformat{equation}{(#2#1#3)}{ and~(#2#1#3)}{, (#2#1#3)}{ and~(#2#1#3)}
\usepackage{subcaption}
\usepackage{commath}
\usepackage{cite}
\usepackage[normalem]{ulem}
\usepackage{textgreek}
\usepackage{soul}
\usepackage{etoolbox}

\usepackage{algorithmic}
\usepackage{algorithm}
\usepackage{textcomp}

\newcommand{\reviewXose}[1]{{\color{red} #1}}
\newcommand{\reviewCommentXose}[1]{{\color{red} #1}}
\newcommand{\minorReviewXose}[1]{{\color{blue} #1}}

\newcommand{\reviewXoseRemoveText}[1]{{\reviewXose{\st{#1}}}}
\newcommand{\minorReviewXoseRemoveText}[1]{{\minorReviewXose{\st{#1}}}}

\newcommand{\secondReviewXose}[1]{{\color{red} #1}}
\newcommand{\secondReviewCommentXose}[1]{{\color{red} #1}}
\newcommand{\secondMinorReviewXose}[1]{{\color{blue} #1}}

\newcommand{\secondReviewXoseRemoveText}[1]{{\secondReviewXose{\st{#1}}}}
\newcommand{\secondMinorReviewXoseRemoveText}[1]{{\secondMinorReviewXose{\st{#1}}}}

\renewcommand{\reviewXose}[1]{{#1}}
\renewcommand{\minorReviewXose}[1]{{#1}}
\renewcommand{\reviewXoseRemoveText}[1]{{}}
\renewcommand{\minorReviewXoseRemoveText}[1]{{}}
\renewcommand{\reviewCommentXose}[1]{{}}

\renewcommand{\secondReviewXose}[1]{{#1}}
\renewcommand{\secondMinorReviewXose}[1]{{#1}}
\renewcommand{\secondReviewXoseRemoveText}[1]{{}}
\renewcommand{\secondMinorReviewXoseRemoveText}[1]{{}}
\renewcommand{\secondReviewCommentXose}[1]{{}}

\begin{document}
	


\bstctlcite{IEEEexample:BSTcontrol}



\def\printBiographies{1} 


	




\acrodef{1G}[1G]{first generation}
\acrodef{2G}[2G]{second generation}
\acrodef{3G}[3G]{third generation}
\acrodef{4G}[4G]{fourth generation}
\acrodef{5G}[5G]{fifth generation}
\acrodef{3GPP}[3GPP]{3rd Generation Partnership Project}
\acrodef{3GPP2}[3GPP2]{3rd Generation Partnership Project 2}
\acrodef{A2A}[A2A]{Air-to-air}
\acrodef{A2G}[A2G]{Air-to-Ground}
\acrodef{ACK}[ACK]{positive acknowledgement}
\acrodef{ADC}[ADC]{analog-to-digital Converter}
\acrodef{ADIF}[ADIF]{Spanish Railway Infrastructure Administrator}
\acrodef{AGC}[AGC]{automatic gain control}
\acrodef{AGCS}[AGCS]{AGC subsystem}
\acrodef{AIC}[AIC]{Akaike's information criteria}
\acrodef{AM}[AM]{Amplitude Modulation}
\acrodef{AMC}[AMC]{Adaptive Modulation and Coding}
\acrodef{AMPS}[AMPS]{Analogue Mobile Phone System}
\acrodef{AoA}[AoA]{angle of arrival}
\acrodef{AP}[AP]{access point}
\acrodef{APp}[AP]{Auxiliary Pilot}
\acrodef{API}[API]{application programming interface}
\acrodef{ARQ}[ARQ]{Automatic Repeat reQuest}
\acrodef{AS}[AS]{Angular Spread}
\acrodef{ASFA}[ASFA]{Automatic Braking and Announcement of Signals}
\acrodef{ASS}[ASS]{Antenna Subset Selection}
\acrodef{ATDMA}[ATDMA]{advanced TDMA mobile access}
\acrodef{AWGN}[AWGN]{additive white Gaussian noise}
\acrodef{BCa}[BCa]{Bias-Corrected and accelerated}
\acrodef{BEM}[BEM]{Basis Expansion Model}
\acrodef{BER}[BER]{bit error ratio}
\acrodef{BIC}[BIC]{Block Interference Cancellation}
\acrodef{BICM}[BICM]{Bit-Interleaved Coded Modulation}
\acrodef{BLER}[BLER]{BLock Error Ratio}
\acrodef{BS}[BS]{Base Station}
\acrodef{BTS}[BTS]{base transceiver station}
\acrodef{BSC}[BSC]{Base Station Controller}
\acrodef{BSS}[BSS]{Base Station Subsystem}
\acrodef{C2}[C2]{communication and control link}
\acrodef{CA}[CA]{Content Aware}
\acrodef{CAP}[CAP]{Coded Auxiliary Pilot}
\acrodef{CCDF}{Complementary Cumulative Distribution Function}
\acrodef{CDF}[CDF]{cumulative distribution function}
\acrodef{CDL}[CDL]{clustered-delay-line}
\acrodef{CDMA}[CDMA]{Code-Division Multiple Access}
\acrodef{CFI}[CFI]{Control Format Indicator}
\acrodef{CIR}[CIR]{channel impulse response}
\acrodef{CL}[CL]{Closed Loop}
\acrodef{CNPC}[CNPC]{control and Non-Payload Communications}
\acrodef{CODIT}[CODIT]{Code Division Testbed}
\acrodef{COFDM}[COFDM]{Coded Orthogonal Frequency Division Multiplexing}
\acrodef{COST}[COST]{European COoperation in the field of Scientific and Technical research}
\acrodef{COTS}[COTS]{Commercial Off-The-Shelf}
\acrodef{CP}[CP]{Cyclic Prefix}
\acrodef{CP-OFDM}[CP-OFDM]{Cyclic Prefix OFDM}
\acrodef{CPICH}[CPICH]{Common Pilot CHannel}
\acrodef{CQI}[CQI]{channel quality indication}
\acrodef{CW}[CW]{continuous wave}
\acrodef{CRC}[CRC]{cyclic redundancy code}
\acrodef{CSI}[CSI]{Channel State Information}
\acrodef{DAC}[DAC]{digital-to-analog converter}
\acrodef{DC}[DC]{Direct Current}
\acrodef{DCCH}[DCCH]{Dedicated Control CHannel}
\acrodef{DCH}[DCH]{Dedicated CHannel}
\acrodef{DCI}[DCI]{downlink control information}
\acrodef{DC}[DC]{direct current}
\acrodef{DDC}[DDC]{Digital Down Converter}
\acrodef{DFT}[DFT]{Discrete Fourier Transform}
\acrodef{DiffServ}[DiffServ]{Differentiate Service}
\acrodef{DMI}[DMI]{Driver-Machine Interface}
\acrodef{DoF}[DoF]{Degrees of Freedom}
\acrodef{DPC}[DPC]{Dirty Paper Coding}
\acrodef{DPCCH}[DPCCH]{Dedicated Physical Control CHannel}
\acrodef{DPDCH}[DPDCH]{Dedicated Physical Data CHannel}
\acrodef{DPS}[DPS]{Discrete Prolate Spheroidal}
\acrodef{DS}[DS]{Double-Stream}
\acrodef{DSP}[DSP]{Digital Signal Processor}
\acrodef{DSTBC}[DSTBC]{Differential Space-Time Block Code}
\acrodef{DSTTDSGRC}[DSTTD-SGRC]{Double Space-Time Transmit Diversity with Sub-Group Rate Control}
\acrodef{DTCH}[DTCH]{Dedicated Traffic CHannel}
\acrodef{DTxAA}[D-TxAA]{Double Transmit Antenna Array}
\acrodef{DUC}[DUC]{Digital Up Converter}
\acrodef{DVB-T}[DVB-T]{Digital Video Broadcasting - Terrestrial}
\acrodef{EAC}[EAC]{EXIT-based Adaptive Coding}
\acrodef{ECR}[ECR]{Effective Code Rate}
\acrodef{EDGE}[EDGE]{Enhanced Data Rates for Global system for mobile communications Evolution}
\acrodef{EESM}[EESM]{Exponential Effective Signal to Interference and Noise Ratio Mapping}
\acrodef{EIRENE}[EIRENE]{European Integrated Railway Radio Enhanced Network}
\acrodef{EIRP}[EIRP]{effective isotropic radiated power}
\acrodef{eNodeB}[eNodeB]{Evolved NodeB}
\acrodef{ePC}[ePC]{Evolved Packet Core}
\acrodef{EPC}[EPC]{Evolved Packet Core}
\acrodef{EPS}[EPS]{Evolved Packet System}
\acrodef{ERTMS}[ERTMS]{European Rail Traffic Management System}
\acrodef{ERTMS/ETCS}[ERTMS/ETCS]{European Rail Traffic Management System/European Train Control System}
\acrodef{ESPRIT}[ESPRIT]{estimation of signal parameters via rotational invariance techniques}
\acrodef{ETCS}[ETCS]{European Train Control System}
\acrodef{ETSI}[ETSI]{European Telecommunications Standards Institute}
\acrodef{eUTRAN}[eUTRAN]{Evolved Universal Terrestrial Radio Access Network}
\acrodef{EVC}[EVC]{European Vital Computer}
\acrodef{EV-DO}[EV-DO]{EVolution-Data Optimized}
\acrodef{EVM}[EVM]{error vector magnitude}
\acrodef{EXIT}[EXIT]{EXtrinsic Information Transfer}
\acrodef{FAA}[FAA]{Federal Aviation Administration}
\acrodef{FBI}[FBI]{FeedBack Information}
\acrodef{FBMC}[FBMC]{Filter Bank Multicarrier}
\acrodef{FDD}[FDD]{frequency-division duplex}
\acrodef{FDTD}[FDTD]{finite-difference time-domain}
\acrodef{FEC}[FEC]{Forward Error Correction}
\acrodef{FFO}[FFO]{fractional frequency offset}
\acrodef{FFT}[FFT]{fast fourier transform}
\acrodef{FLoS}[FLoS]{Far Line-of-Sight}
\acrodef{FM}[FM]{Frequency Modulated}
\acrodef{FMT}[FMT]{Filtered MultiTone}
\acrodef{fOFDM}[fOFDM]{Filtered OFDM}
\acrodef{FPGA}[FPGA]{field programmable gate array}
\acrodef{FRS}[FRS]{Functional Requirement Specification}
\acrodef{FTP}[FTP]{File Transfer Protocol}
\acrodef{GFDM}[GFDM]{Generalized FDM}
\acrodef{GGSN}[GGSN]{Gateway-General packet radio service Support Node}
\acrodef{GO}[GO]{geometric optics}
\acrodef{GOF}[GOF]{Goodness Of Fit}
\acrodef{GOP}[GOP]{Group Of Pictures}
\acrodef{GPRS}[GPRS]{General Packet Radio Service}
\acrodef{GPS}[GPS]{Global Positioning System}
\acrodef{GPU}[GPU]{Graphics Processing Unit}
\acrodef{GSM}[GSM]{Global System for Mobile Communications}
\acrodef{GSM-R}[GSM-R]{GSM for Railways}
\acrodef{GSTM}[GSTM]{Geometry-Based Spatial-Consistent MPC Tracking Method}
\acrodef{GTIS}[GTIS]{GTEC Testbed Interface Software}
\acrodef{HARQ}[HARQ]{hybrid automatic repeat request}
\acrodef{HDTV}[HDTV]{High Definition TV}
\acrodef{HI}[HI]{HARQ indicator}
\acrodef{HRPE}[HRPE]{High-Resolution Parameter Estimation}
\acrodef{HSDPA}[HSDPA]{High Speed Downlink Packet Access}
\acrodef{HSDPCCH}[HS-DPCCH]{High-Speed Dedicated Physical Control CHannel}
\acrodef{HSDSCH}[HS-DSCH]{High-Speed Downlink Shared CHannel}
\acrodef{HSPA}[HSPA]{High-Speed Packet Access}
\acrodef{HSPDSCH}[HS-PDSCH]{High-Speed Physical Downlink Shared CHannel}
\acrodef{HSSCCH}[HS-SCCH]{High-Speed Shared Control CHannel}
\acrodef{HST}[HST]{high-speed train}
\acrodef{HSUPA}[HSUPA]{High-Speed Uplink Packet Access}
\acrodef{IBx}[IBx]{Indoor Office B}
\acrodef{IC}[IC]{Interference Cancellation}
\acrodef{ICI}[ICI]{inter-carrier interference}
\acrodef{IDE}[IDE]{Integrated Development Environment}
\acrodef{IEEE}[IEEE]{Institute of Electrical and Electronics Engineers}
\acrodef{IF}[IF]{intermediate frequency}
\acrodef{IFFT}[IFFT]{inverse fast fourier transform}
\acrodef{IFO}[IFO]{integer frequency offset}
\acrodef{iid}[iid]{independent and identically-distributed}
\acrodef{IMPC}[IMPC]{Instantaneous MPC estimate}
\acrodef{IMT}[IMT]{International Mobile Telecommunications}
\acrodef{IMT-A}[IMT-A]{IMT-Advanced}
\acrodef{IP}[IP]{Internet Protocol}
\acrodef{IR}[IR]{Incremental Redundancy}
\acrodef{ISI}[ISI]{Inter-Symbol Interference}
\acrodef{ISM}[ISM]{industrial, scientific and medical}
\acrodef{ISO}[ISO]{International Standard Organization}
\acrodef{ITU}[ITU]{International Telecommunication Union}
\acrodef{ITU-R}[ITU-R]{ITU Radiocommunication Sector}
\acrodef{JADE}[JADE]{Joint Approximate Diagonalization of Eigenmatrices}
\acrodef{J-TACS}[J-TACS]{Japanese Total Access Communication System}
\acrodef{KEST}[KEST]{Kalman Enhanced Super Resolution Tracking}
\acrodef{KL}[KL]{Karhunen-Loève}
\acrodef{KLT}[KLT]{Karhunen-Loève Transform}
\acrodef{KP}[KP]{kilometric point}
\acrodef{LAN}[LAN]{local area network}
\acrodef{LDGM}[LDGM]{Low Density Generator Matrix}
\acrodef{LDPC}[LDPC]{Low Density Parity Check}
\acrodef{LEU}[LEU]{Lineside Electronic Units}
\acrodef{LMMSE}{Linear Minimum Mean Square Error}
\acrodef{LMPC}[LMPC]{Long-term Time-Variant MPC}
\acrodef{LNA}[LNA]{Low-Noise Amplifier}
\acrodef{LOGS}[LOGS]{log subsystem}
\acrodef{LoS}[LoS]{line-of-sight}
\acrodef{LS}[LS]{least squares}
\acrodef{LTE}[LTE]{Long Term Evolution}
\acrodef{LTE-A}[LTE-A]{LTE-Advanced}
\acrodef{MAC}[MAC]{Medium Access Control}
\acrodef{MACd}[MAC-d]{Medium Access Control dedicated}
\acrodef{MAChs}[MAC-hs]{Medium Access Control for High-Speed Downlink Packet Access}
\acrodef{maxCI}[max C/I]{Maximum Carrier-to-Interference Ratio}
\acrodef{MAMOS}[MAMOS]{management and monitoring subsystem}
\acrodef{MAP}{Maximum A-Posteriori}
\acrodef{MB}[MB]{MacroBlock}
\acrodef{MCS}{Modulation and Coding Scheme}
\acrodef{MCM}[MCM]{multi carrier modulation}
\acrodef{MF}[MF]{Matched Filter}
\acrodef{MI}[MI]{Mutual Information}
\acrodef{MIESM}[MIESM]{Mutual Information Effective Signal to Interference and Noise Ratio Mapping}
\acrodef{MIMO}[MIMO]{multiple-input, multiple-output}
\acrodef{MISO}[MISO]{Multiple-Input Single-Output}
\acrodef{ML}[ML]{maximum likelihood}
\acrodef{ML-EM}[ML-EM]{Maximum Likelihood Expectation-Maximization}
\acrodef{MMSE}[MMSE]{Minimum Mean Squared Error}
\acrodef{mmWVs}[mmWVs]{millimeter-waves}
\acrodef{MORANE}[MORANE]{Mobile Radio for Railways Networks in Europe}
\acrodef{MoRSE}[MoRSE]{Mobile Radio Simulation Environment}
\acrodef{MPC}[MPC]{Multipath Component}
\acrodef{MPEG}[MPEG]{Moving Picture Expert Group}
\acrodef{MRC}[MRC]{Maximum Ratio Combining}
\acrodef{MRCN}[MRCN]{Mobile Radio Center}
\acrodef{MS}[MS]{Mobile Station}
\acrodef{MSC}[MSC]{Mobile Switching Center}
\acrodef{MSE}[MSE]{mean square error}
\acrodef{MTU}[MTU]{Maximum Transfer Unit}
\acrodef{MU}[MU]{Multi-User}
\acrodef{MUSIC}[MUSIC]{multiple signal classification}
\acrodef{MVC}[MVC]{Model View Controller}
\acrodef{MVU}[MVU]{Minimum Variance Unbiased}
\acrodef{MWL}[MWL]{Middleware Layer}
\acrodef{NACK}[NACK]{negative acknowledgement}
\acrodef{NAL}[NAL]{Network Abstract Layer}
\acrodef{NASA}[NASA]{National Aeronautics and Space Administration}
\acrodef{NBAP}[NBAP]{NodeB Application Part}
\acrodef{NDI}[NDI]{New Data Indicator}
\acrodef{NGMN}[NGMN]{Next Generation Mobile Networks}
\acrodef{NHSIIT}[NHSIIT]{high speed integrated inspection train}
\acrodef{NLoS}[NLoS]{non-line-of-sight}
\acrodef{NMT}[NMT]{Nordic Mobile Telephone}
\acrodef{NSAW}[N-SAW]{N Stop And Wait}
\acrodef{NSN}[NSN]{Nokia Siemens Networks}
\acrodef{NSS}[NSS]{Network Switching Subsystem}
\acrodef{OFDM}[OFDM]{orthogonal frequency-division multiplexing}
\acrodef{OFDMA}[OFDMA]{Orthogonal Frequency-Division Multiple Access}
\acrodef{OLoS}[OLoS]{Obstructed LoS}
\acrodef{OPTA}[OPTA]{Optimal Performance Theoretically Attainable}
\acrodef{OQAM}[OQAM]{offset quadrature amplitude modulation}
\acrodef{OSTBC}[OSTBC]{Orthogonal Space-Time Block Code}
\acrodef{OSI}[OSI]{Open Systems Interconnection}
\acrodef{PA}[PA]{Public Address}
\acrodef{PAPR}[PAPR]{Peak-to-Average Power Ratio}
\acrodef{PARC}[PARC]{Per-Antenna Rate Control}
\acrodef{PBCH}[PBCH]{Physical Broadcast Channel}
\acrodef{PC}[PC]{personal computer}
\acrodef{PCA}[PCA]{Principal Component Analysis}
\acrodef{PCI}[PCI]{Physical Cell Identity}
\acrodef{PCIe}[PCIe]{Peripheral Component Interconnect Express}
\acrodef{PCFICH}[PCFICH]{Physical Control Format Indicator Channel}
\acrodef{PDCCH}[PDCCH]{Physical Downlink Control Channel}
\acrodef{PDCP}[PDCP]{Packet Data Convergence Protocol}
\acrodef{pdf}[pdf]{probability density function}
\acrodef{PDP}[PDP]{power delay profile}
\acrodef{PDSCH}[PDSCH]{Physical Downlink Shared Channel}
\acrodef{PDU}[PDU]{Packet Data Unit}
\acrodef{PedA}[PedA]{Pedestrian A}
\acrodef{PedB}[PedB]{Pedestrian B}
\acrodef{PER}[PER]{Packet Error Rate}
\acrodef{PERS}[PERS]{persistence subsystem}
\acrodef{PF}[PF]{Proportional Fair}
\acrodef{PHICH}[PHICH]{Physical hybrid-ARQ Indicator Channel}
\acrodef{PN}[PN]{Pseudo-Noise}
\acrodef{PPS}[PPS]{Pulse Per Second}
\acrodef{PRCN}[PRCN]{Portable Radio Center}
\acrodef{PSD}[PSD]{power spectral density}
\acrodef{PSK}[PSK]{Phase Shift Keying}
\acrodef{PSNR}[PSNR]{Peak Signal to Noise Ratio}
\acrodef{P-SCH}[P-SCH]{Primary Synchronization Signal}
\acrodef{QAM}[QAM]{quadrature amplitude modulation}
\acrodef{QCIF}[QCIF]{Quarter Common Intermediate Format}
\acrodef{QoE}[QoE]{Quality of Experience}
\acrodef{QoS}[QoS]{Quality of Service}
\acrodef{RAID}[RAID]{Redundant Array of Independent Disks}
\acrodef{RAN}[RAN]{Radio Access Network}
\acrodef{RAx}[RAx]{Rural Area channel model}
\acrodef{RBC}[RBC]{Radio Bearer Control}
\acrodef{RBCN}[RBCN]{Radio Block Center}
\acrodef{RCC}[RCC]{Radio Control Center}
\acrodef{RENFE}[RENFE]{Red Nacional de los Ferrocarriles Espa\~{n}oles}
\acrodef{RMS}[RMS]{root mean square}
\acrodef{RF}[RF]{radio frequency}
\acrodef{RLC}[RLC]{Radio Link Control}
\acrodef{RNC}[RNC]{Radio Network Controller}
\acrodef{RoC}[RoC]{Radius of curvature}
\acrodef{RR}[RR]{Round Robin}
\acrodef{RRC}[RRC]{Radio Resource Control}
\acrodef{RRM}[RRM]{Radio Resource Management}
\acrodef{RSSI}[RSSI]{received signal strength indication}
\acrodef{RTP}[RTP]{Real-time Transport Protocol}
\acrodef{RV}[RV]{Redundancy Version}
\acrodef{RX}[RX]{Receiver}
\acrodef{SAE}[SAE]{System Architecture Evolution}
\acrodef{SAGE}[SAGE]{space-alternating generalized expectation-maximization}
\acrodef{SC-FDMA}[SC-FDMA]{Single Carrier Frequency-Division Multiple Access}
\acrodef{SCLDGM}[SCLDGM]{Serially-Concatenated Low Density Generator Matrix}
\acrodef{SCM}[SCM]{spatial channel model}
\acrodef{SCME}[SCME]{SCM-Extension}
\acrodef{SDMA}[SDMA]{Spatial Division Multiple Access}
\acrodef{SDR}[SDR]{Software-Defined Radio}
\acrodef{SDU}[SDU]{Service Data Unit}
\acrodef{SER}[SER]{Symbol Error Ratio}
\acrodef{SFDR}[SFDR]{Spurious-Free Dynamic Range}
\acrodef{SGSN}[SGSN]{Serving-General packet radio service Support Node}
\acrodef{SHB}[SHB]{Sundance High-Speed Bus}
\acrodef{SIC}[SIC]{Successive Interference Cancellation}
\acrodef{SID}[SID]{Size Index Identifier}
\acrodef{SIM}[SIM]{Subscriber Identity Module}
\acrodef{SIMO}[SIMO]{Single-Input Multiple-Output}
\acrodef{SINR}[SINR]{Signal to Interference and Noise Ratio}
\acrodef{SIR}{Signal-to-Interference Ratio}
\acrodef{SIS}[SIS]{signal integrity subsystem}
\acrodef{SMPC}[SMPC]{Short-term Time-Variant MPC}
\acrodef{SMT}[SMT]{Staggered Multitone}
\acrodef{SNCF}[SNCF]{Soci\'et\'e Nationale des Chemins de Fer Fran\c{c}ais}
\acrodef{SNR}[SNR]{Signal to Noise Ratio}
\acrodef{SOS}[SOS]{Second-Order Statistics}
\acrodef{SPL}[SPL]{Signal Processing Layer}
\acrodef{SQP}[SQP]{Sequential Quadratic Programming}
\acrodef{SRS}[SRS]{System Requirement Specification}
\acrodef{SS}[SS]{Single-Stream}
\acrodef{SISO}[SISO]{Single-Input Single-Output}
\acrodef{SSD}[SSD]{solid-state drive}
\acrodef{STBC}[STBC]{Space-Time Block Code}
\acrodef{STMMSE}[ST-MMSE]{Space-Time Minimum Mean Squared Error}
\acrodef{STTD}[STTD]{Space-Time Transmit Diversity}
\acrodef{SU}[SU]{Single-User}
\acrodef{SVD}[SVD]{Singular Value Decomposition}
\acrodef{S-SCH}[S-SCH]{Secondary Synchronization Signal}
\acrodef{S-SCH1}[${\textrm{S-SCH}}_1$]{First Secondary Synchronization Sequence}
\acrodef{S-SCH2}[${\textrm{S-SCH}}_2$]{Second Secondary Synchronization Sequence}
\acrodef{TACS}[TACS]{Total Access Communication System}
\acrodef{TB}[TB]{transport block}
\acrodef{TBM}[TBM]{Tunnel Boring Machine}
\acrodef{TBS}[TBS]{Transport Block Size}
\acrodef{TCP}[TCP]{Transmission Control Protocol}
\acrodef{TDD}[TDD]{time-division duplex}
\acrodef{TD-LTE}[TD-LTE]{Time-Division LTE}
\acrodef{TFC}[TFC]{Transport Format Combination}
\acrodef{TFCI}[TFCI]{Transport-Format Combination Indicator}
\acrodef{TFT}[TFT]{Traffic Flow Template}
\acrodef{TIL}[TIL]{Testbed Interface Layer}
\acrodef{TOS}[TOS]{Type Of Service}
\acrodef{TPC}[TPC]{Transmit Power-Control}
\acrodef{TrCH}[TrCH]{Transport CHannel}
\acrodef{TSN}[TSN]{Transmission Sequence Number}
\acrodef{TTI}[TTI]{Transmission Time Interval}
\acrodef{TUx}[TUx]{Typical Urban channel model}
\acrodef{TX}[TX]{Transmitter}
\acrodef{TxAA}[TxAA]{Transmit Antenna Array}
\acrodef{UAV}[UAV]{Unmanned Aerial Vehicle}
\acrodef{UDP}[UDP]{User Datagram Protocol}
\acrodef{UE}[UE]{user equipment}
\acrodef{UHD}[UHD]{USRP hardware driver}
\acrodef{UHDIS}[UHDIS]{UHD interface subsystem}
\acrodef{UHF}[UHF]{ultra high frequency}
\acrodef{UIC}[UIC]{Union Internationale des Chemins de Fer}
\acrodef{UMTS}[UMTS]{Universal Mobile Telecommunications System}
\acrodef{UPS}[UPS]{uninterruptible power supply}
\acrodef{USB}[USB]{Universal Serial Bus}
\acrodef{USRP}[USRP]{Universal Software Radio Peripheral}
\acrodef{UTRAN}[UTRAN]{Universal mobile telecommunications system Terrestrial Radio Access Network}
\acrodef{UWB}[UWB]{Ultra Wide-Band}
\acrodef{V2I}[V2I]{Vehicle-to-Infrastructure}
\acrodef{V2V}[V2V]{Vehicle-to-Vehicle}
\acrodef{VBLAST}[V-BLAST]{Vertical Bell Laboratories Layered Space-Time}
\acrodef{VCEG}[VCEG]{Video Coding Expert Group}
\acrodef{VCL}[VCL]{Video Coding Layer}
\acrodef{VehA}[VehA]{Vehicular A}
\acrodef{VHF}[VHF]{very high frequency}
\acrodef{VHSIC}[VHSIC]{Very High Speed Integrated Circuits}
\acrodef{VLoS}[VLoS]{Visual Line-of-Sight}
\acrodef{VoD}[VoD]{Video on Demand}
\acrodef{VoIP}[VoIP]{Voice over Internet Protocol}
\acrodef{WCDMA}[WCDMA]{Wideband Code Division Multiple Access}
\acrodef{WiFi}[WiFi]{Wireless Fidelity}
\acrodef{WiMAX}[WiMAX]{Worldwide Interoperability for Microwave Access}
\acrodef{WLAN}[WLAN]{wireless local area network}
\acrodef{WSS}[WSS]{Wide Sense Stationary}
\acrodef{WSSUS}[WSSUS]{wide-sense stationary uncorrelated scattering}
\acrodef{ZF}[ZF]{Zero-Forcing}







\title{Air-to-Ground Channel Characterization for Low-Height UAVs in Realistic Network Deployments}



\author{Jos\'e~Rodr\'iguez-Pi\~neiro,~
	Tom\'as~Dom\'inguez-Bola\~no,~
	Xuesong~Cai,~
	Zeyu~Huang,~
	and~Xuefeng~Yin,~\IEEEmembership{Member,~IEEE}
	\thanks{J.~Rodr\'iguez-Pi\~neiro and Z.~Huang are with the College of Electronics and Information Engineering, Tongji University, Shanghai, China. (e-mail: j.rpineiro@tongji.edu.cn and huangzeyu@tongji.edu.cn).}
	\thanks{T.~Dom\'inguez-Bola\~no is with the Department of Computer Engineering and the CITIC Research Center, University of A Coru\~na, A Coru\~na, Spain and the. (e-mail: tomas.bolano@udc.es).}
	\thanks{X.~Cai is with the Wireless Communication Networks Section, Department of Electronic Systems, Aalborg University, Aalborg, Denmark. (e-mail: xuc@es.aau.dk).}
	\thanks{X.~Yin is with the College of Electronics and Information Engineering and the National Computer and Information Technology Practical Education Demonstration Center, Tongji University, Shanghai, China. (e-mail: yinxuefeng@tongji.edu.cn).}
	\thanks{Corresponding author: Xuesong~Cai (e-mail address: xuc@es.aau.dk).}
	\thanks{{This work was supported by the
		National Natural Science Foundation of China (NSFC) under Grants 61850410529 and 61971313; as well as by the Xunta de Galicia (ED431G2019/01), the Agencia Estatal de Investigaci\'on of Spain (TEC2016-75067-C4-1-R, PID2019-104958RB-C42) and ERDF funds of the EU (AEI/FEDER, UE). The authors also want to specially thank Prof. Preben E. Mogensen for his support in publishing this paper.}}}


\maketitle


\begin{abstract}
	Due to the decrease in cost, size and weight, \acp{UAV} are becoming more and more popular for general-purpose civil and commercial applications. Provision of communication services to \acp{UAV} both for user data and control messaging by using off-the-shelf terrestrial cellular deployments introduces several technical challenges. In this paper, an approach to the air-to-ground channel characterization for low-height \acp{UAV} based on an extensive measurement campaign is proposed, giving special attention to the comparison of the results when a typical directional antenna for network deployments is used and when a quasi-omnidirectional one is considered. Channel characteristics like path loss, shadow fading, root mean square delay and Doppler frequency spreads and the K-factor are statistically characterized for different suburban scenarios.
\end{abstract}

\begin{IEEEkeywords}
	UAV; Air-to-Ground; Communications channels; Time-varying channels; Aircraft communication
\end{IEEEkeywords}

\acresetall



\section{Introduction}


Small \acp{UAV} are rapidly changing their main scope from the traditional military usage in hostile environments \cite{zeng2016wireless,valavanis2014handbook} to general-purpose civil and commercial applications. The decrease in their cost, size and weight, the increase of their battery life, their high maneuverability and their ability to hover \cite{hayat2016survey} make them an appropriate tool for a wide set of applications, such as border surveillance, operations in inaccessible areas, delivery of goods, search and rescue missions \cite{hayat2016survey,ryan2004overview,degarmo2004issues,kovacina2002multi}, precise land mapping by aerial imagery \cite{ArtigoReconstrucionPlanialtimetrica_ICA-ACCA_2018,ArtigoSistemaDeTransicionDeVoo_ICA-ACCA_2018,tariq2016development,segales2016implementation} or precise farming \cite{us2013unmanned,zhang2012application}. The provision of temporary network access after natural disasters, emergency situations or in saturated environments became one of the key scenarios addressed by the \ac{5G} communication systems \cite{zeng2016wireless,osseiran2014scenarios} due to the ability of the \acp{UAV} for fast deployments \cite{zhan2011wireless,zhou2015multi,lee2011role,palat2005cooperative,gu2000uav,valcarce2013airborne}.

Regardless of the application under consideration, \ac{UAV} communications can be classified into two types:

\begin{itemize}
	\item\textbf{Payload-oriented communications}: used to transmit non-critical user data, in general they seek to maximize the data rate in a best-effort manner, being tolerant to errors or delays. This could be the case e.g. of transmitting the video signal from the on-board cameras in digital imagery.
	\item\textbf{Critical communications}: they involve safety and control-related messages and usually imply low data rate requirements, but very high \ac{QoS} standards in terms of delay and availability. Although in many cases \acp{UAV} could fly autonomously, it may be required to reliably change some settings during operation for safety purposes. This is one prerequisite for \ac{UAV} traffic management, an area under exploration by the \ac{FAA} and the \ac{NASA} \cite{lin2017sky,yanmaz2013achieving}.
\end{itemize}

The use of ground \ac{LTE} or future \ac{5G} deployments to provide network access to \acp{UAV} devoted a great level of interest by the research community. In fact, the \ac{3GPP} approved the corresponding study and work items \cite{LTEUAVSI,LTEUAVWI} to investigate the feasibility of serving \acp{UAV} by terrestrial \ac{LTE} deployments \cite{ArtigoModeladoCanleUAV2017}. However, the use of a terrestrial deployment for the provision of network access to \acp{UAV} proposes a set of major signal propagation challenges regarding coverage and interference management. In these deployments the antennas are focused on serving users at ground level and thus they are down-tilted. This way, the ground itself, as well as the terrestrial elements such as buildings, define the geometric area for each cell and \minorReviewXose{limit} its interference in the neighboring cells. In the case of \acp{UAV}, for public-safety reasons, most of the countries limit the applications to low altitude flights (below $150$\,m) under \ac{VLoS} conditions \cite{european2013roadmap,amorim2017radio,ArtigoModeladoCanleUAV2017}. On the one side, the down-tilted configuration of the \ac{BS} antennas is not optimum for serving \acp{UAV}, even for low-height applications. On the other side, even if the configuration of the \ac{BS} antennas was adapted to provide sky coverage, \reviewXose{the strategies for management of the inter-cell interference would probably need to be reviewed}, since the propagation is naturally more \ac{LoS}-like \cite{welch2016evolving} and no obstacles would help to limit each cell, differently from the terrestrial case \cite{van2016lte}. Moreover, \reviewXose{this would be also true for the uplink communications (i.e., those from the \ac{UAV} to the \ac{BS}), since the area of interference of an \ac{UAV} would be less limited by the ground elements and probably affect both to the neighboring cells as well as to other potential nearby \acp{UAV}, hence impacting both terrestrial and aerial users \cite{lin2017sky}. Indeed, the interference is one of the main focus of the \ac{3GPP} study item on enhanced \ac{LTE} support for aerial vehicles \cite{ArtigoModeladoCanleUAV2017,lin2017sky,ntt2017study}}. More challenges from the propagation point of view arise from the influence of the \ac{UAV} structure itself in the shadowing characteristics \cite{matolak2014initial} or due to the fact that, since the flights are performed at low altitudes, the ground environment has non-negligible effects on the communications channel. \reviewXose{Due to the propagation challenges mentioned, the accurate study of the \ac{A2G} communications channel for low-height \acp{UAV} from a propagation point of view becomes essential. An accurate statistical characterization of the communications channel between a \ac{BS} and an \ac{UAV} is the first step to effectively evaluate different strategies for interference mitigation or network deployments in complex environments consisting of several \acp{BS} --with either directional/sectorial or omnidirectional radiation patterns-- and users by using tools such as system-level simulators.}

\subsection{Related Work}\label{sec:relatedWork}

Different approaches to the characterization of the \ac{A2G} communications channel for \acp{UAV} are proposed in the literature, based on simulations and/or measurements. In the next paragraphs the main related works are analyzed. 

\subsubsection{\textbf{A2G Channel Characterization by Simulations}}

In \cite{lin2017sky}, the authors propose the use of the \ac{3GPP} channel models in \cite{3gpp2017study} for \acp{UAV} at altitudes below the base station antenna height and free-space propagation for higher cases. In \cite{feng2006path} new statistical models for \ac{A2G} channels in the range of frequencies between $200$\,MHz and $5$\,GHz and urban environments are provided. In \cite{tameh19973} the influence of the elevation angle on the path loss and shadowing are evaluated from the simulated propagation data extracted from a three-dimensional outdoor deterministic ray-tracing model. Ray-tracing is also used in \cite{khawaja2017uav} to characterize mmWave propagation ($28$\,GHz and $60$\,GHz) for urban, suburban, rural and over sea environments.

\subsubsection{\textbf{A2G Channel Characterization by Measurements}}


Several works focused on the characterization of the path loss for the \ac{A2G} channel for \acp{UAV}. 
In \cite{al2017modeling}, the authors obtain the aerial path loss as an excess value to the path loss that would correspond to a terrestrial user for a \ac{LTE} cellular deployment in the band of $850$\,MHz in typical suburban environments. Flight heights between $15$\,m and $120$\,m were considered. In \cite{yanmaz2011channel}, modeling of the path loss exponents for the \ac{A2G} link in open field and campus scenarios is considered. Height values between $20$\,m and $120$\,m at $100$\,m ground distance between the transmitter and the receiver are considered. The effect of the \ac{UAV} orientation is also studied. 

In \cite{amorim2017radio}, the shadowing is also characterized, along with the  path loss. The authors consider the radio channel between \acp{UAV} and commercial \ac{LTE} \acp{BS} at the $800$\,MHz band, link distances between $1$\,km and $22$\,km, and flight heights between $1.5$\,m and $120$\,m. In \cite{khawaja2016uwb}, statistical models to characterize not only the large-scale fading but also the small-scale fading and multipath propagation are proposed. The work is based on a \ac{UWB} measurement campaign in the frequency range between $3.1$\,GHz and $5.3$\,GHz on several scenarios including blockage of the on-ground receiver by foliage or not and with very low flight heights between $4$\,m and $16$\,m. The multipath propagation is characterized by means of the the \ac{PDP} and the \ac{RMS} delay spread.

In \cite{goddemeier2012role}, a height and distance-aware aerial radio channel model is derived from measurements taken with a helium balloon in stationary positions at heights up to $500$\,m. The ground distance between the base station and the receiver is $1900$\,m. The tests are performed by passive sounding of \ac{GSM} and \ac{UMTS} signals in an urban environment at a center frequency of $2120$\,MHz.

In \cite{izydorczyk2019angular} the variation of the mean \ac{AoA} and \ac{AS} with flight height is evaluated based on experimental measurements in both urban and rural scenarios using commercial \ac{LTE} deployments. Since the authors consider a large antenna array, they used a crane to lift it to different heights, instead of an actual \ac{UAV} during the measurements. By using a similar deployment, in \cite{izydorczyk2018experimental} the authors study the performance of different multi-antenna receiver techniques for \ac{UAV} communications.

Probably the richest set of recent measurements regarding \ac{A2G} channel characterization for non-low \ac{UAV} flight heights is provided by D. Matolak, \textit{et al.} in \cite{matolak2015air,matolak2014air,matolak2016air,sun2017air,matolak2017air,matolak2017air2,matolak2014initial,matolak2013air,matolak2015unmanned,matolak2014air2,matolak2012air}, among others. Sections of the L and C bands \cite{ITUR4318} are considered, with bandwidths of $5$\,MHz and $50$\,MHz, respectively. As the ground transmitter, a transportable tower with variable height (from $4$\,m to $20$\,m) was considered, whereas a piloted aircraft was used as the receiver, being the flight height values between $500$\,m and $2000$\,m and the link distances between $500$\,m to $50$\,km, approximately. Channel parameters such as path loss, delay spread, stationarity distance, K-factor or inter-band and spatial correlation were evaluated. Characterizations for path loss and \acp{MPC} were proposed for different environments, such as over-water (see \cite{matolak2017air,matolak2014initial}) or over-freshwater (see \cite{matolak2014air2}), hilly and mountainous environments (see \cite{sun2017air}), suburban or near-urban scenarios (see \cite{matolak2017air2,matolak2015air}), and hilly suburban environments (see \cite{matolak2014air}). In \cite{schneckenburger2017modeling}, a geometrical-statistical channel modeling approach for the \ac{A2G} channel in L-band is considered, whereas the authors show how the channel parameters can be derived from the measurement data in \cite{matolak2017air,schneckenburger2016measurement}. In the latter work, results from flight trials with an aircraft for characterizing the \ac{A2G} channel for the L-band in positioning applications are shown, being the considered bandwidth $10$\,MHz. The work, being an extension of \cite{schneckenburger2014band} considers \ac{PDP}, Doppler frequency Delay Profile, mean delay, \ac{RMS} Doppler frequency spread and \ac{RMS} delay spread, as well as ranging accuracy results. Different scenarios such as in-route cruise, climb-and-descent and takeoff-and-landing are considered. Flight altitudes range from $3$\,km to $9$\,km, except for the takeoff-and landing scenario, which considers heights between $30$\,m and $330$\,m. Link distance ranges from $500$\,m to $350$\,km.

Based on a extensive measurement campaign using an \ac{UAV} and a commercial \ac{LTE} \ac{BS} in suburban environments, the publications \cite{ArtigoModeladoCanleUAV2017,ArtigoMedidasPasivasUav_EUCAP2019,ArtigoModeladoUavintelixenciaArtificial_GLOBECOM2017} were released. In \cite{ArtigoModeladoCanleUAV2017,ArtigoMedidasPasivasUav_EUCAP2019} a stochastic channel model was proposed,  including characterizations of path loss, shadow fading, delay spread and Doppler frequency spread for horizontal and vertical flights at different heights and distances to the \ac{BS}. In \cite{ArtigoModeladoUavintelixenciaArtificial_GLOBECOM2017}, a big-data-assisted channel modeling strategy is applied to find the most sensitive channel parameter from a specific set for the \ac{A2G} \ac{UAV} channel and to characterize it (the selected parameter was the K-factor). In \cite{ArtigoGraphModellingUAV_EUCAP2019}, some of the measurement environments in the current paper are considered to test the accuracy of graph modeling channel simulation techniques to reconstruct the \ac{PDP} and \acp{MPC} of the \ac{A2G} \ac{UAV} channel. It is also worth noting that \cite{ArtigoAvaliacionUAVsConCamaraAnecoica_GLOBECOM2018} proposes channel emulation in multi-probe anechoic chambers as a feasible alternative to extensive measurement campaigns.

Increasing interest in characterizing \ac{MIMO} channels technologies for \ac{A2G} communications is also appreciated. E.g., in \cite{wentz2015mimo}, the GAGE channel model proposed in \cite{newhall2002geometric}, further developed in \cite{newhall2003radio} by means of the measurement results obtained in \cite{newhall2003wideband}, was extended to the \ac{MIMO} case. Test measurements of \ac{MIMO} communications for the \ac{A2G} channels are reported in \cite{kung2010measuring} ($3\times 4$ \ac{MIMO}) or \cite{chen2011mimo} ($4\times 4$ case). In \cite{willink2016measurement}, the communications channel between a 8-antennas ground receiver and a two-antennas transmitter low-altitude \ac{UAV} is studied by means of measurements. Temporal and spatial properties of the channel are studied for flight heights of approximately $200$\,m, for a carrier frequency of $915$\,MHz, and a bandwidth of $10$\,MHz and two different distances between the receiver and the flight route.

\subsection{Main Contributions}

\begin{itemize}
	\item In this manuscript, a measurement campaign for the \ac{A2G} low-height \ac{UAV} communications channel was performed. The measurement campaign  systematically studied different suburban environments at different flight height values between $15$\,m and $105$\,m. Furthermore, both an omnidirectional antenna as well as typical \ac{BS} one were considered at the transmitter. While the results for the omnidirectional antenna allow for obtaining an accurate characterization of the channel characteristics, results for the directional \ac{BS} antenna show the behavior \reviewXose{for an individual sector of a \ac{BS} (note that, in practical deployments, several sectors per \ac{BS} are set up)}.
	\item A \ac{HRPE} algorithm was used to estimate the attenuation, delay and Doppler frequency values for the different \acp{MPC} exhibited by the channel. Based on those results, the communications channel was statistically characterized in terms of the path loss, shadow fading, \ac{RMS} delay and Doppler frequency spreads and K-factor. The variation of the mentioned channel characteristics with the flight height, distance between the \ac{UAV} and the \ac{BS}, the ground environment and the considered antenna at the \ac{BS} was systematically studied. The obtained channel characteristics provide valuable insights for the design of \ac{UAV} communications systems and their applications, as well as for the related network deployments.
\end{itemize}

The structure of this paper is as follows: \cref{sec:MeasurementEnvironmentAndProcedure} describes the measurement environment and setup considered, whereas \cref{sec:SignalGenerationAndProcessing} describes how the acquired samples were processed. \cref{sec:ChannelCharacterizationForTheFlights} statistically characterizes the channel observed for the different measured environments \reviewXose{and} transmitter configurations, and \cref{sec:Conclusions} contains the main conclusions of the performed study.

\section{Measurements Environment and Procedure}\label{sec:MeasurementEnvironmentAndProcedure}

In this section, a systematic measurement for characterizing the \ac{A2G} low-height \ac{UAV} communications channel is detailed. \cref{sec:MeasurementEnvironmentDetails} defines the considered measurement environments, whereas \cref{sec:MeasurementEquipmentDetails} describes the employed measurement setup.

\subsection{Measurement Environment}\label{sec:MeasurementEnvironmentDetails}

\begin{figure}[t!]
	\centering
	\begin{subfigure}[t]{\linewidth}
		\centering
		\includegraphics[width=\columnwidth]{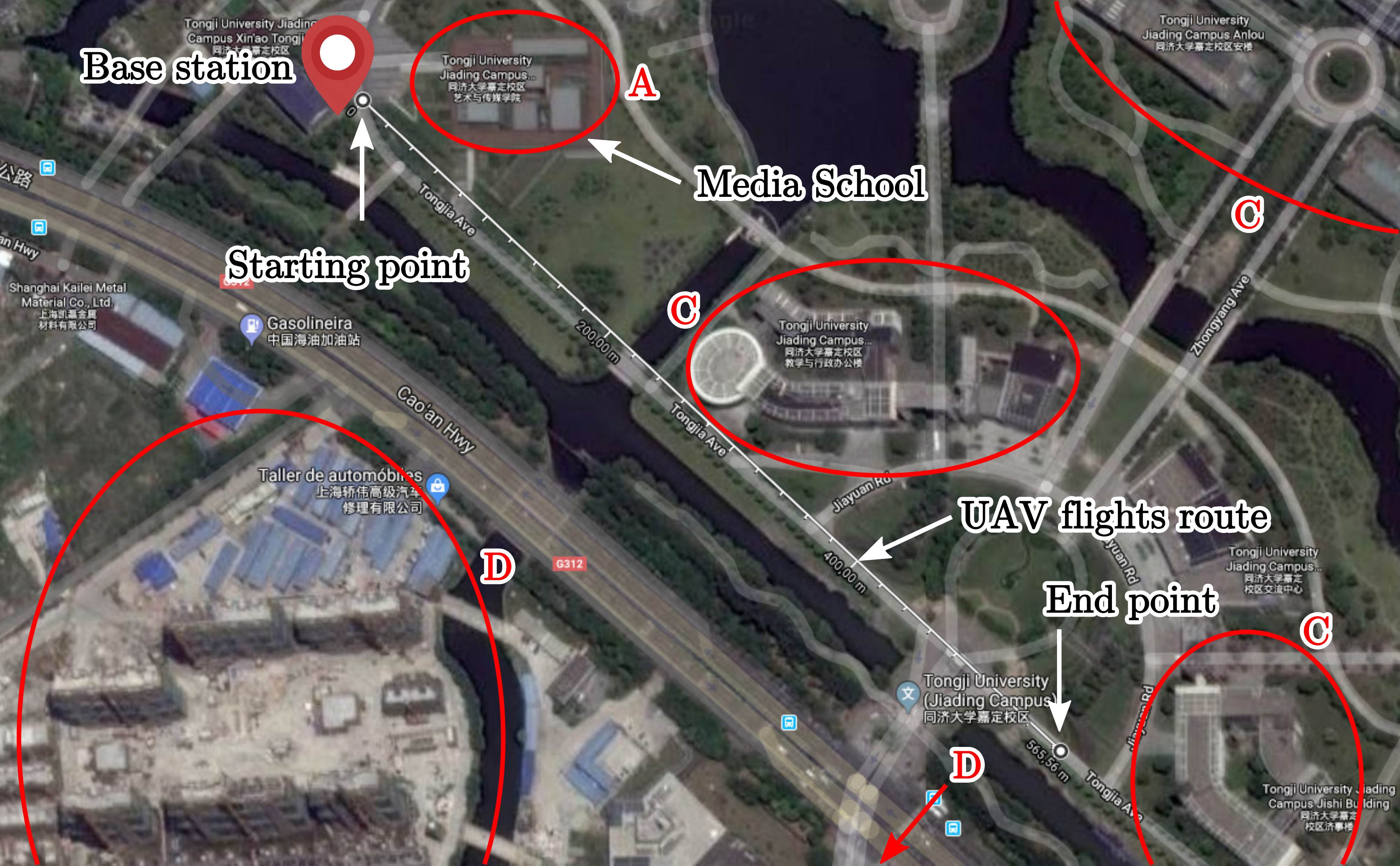}
		\caption{Measurement Environment~I (\ac{LoS}). \secondReviewCommentXose{This figure was modified.}}
		\label{fig:los}
	\end{subfigure}
	\par\medskip
	\begin{subfigure}[t]{\linewidth}
		\centering
		\includegraphics[width=\columnwidth]{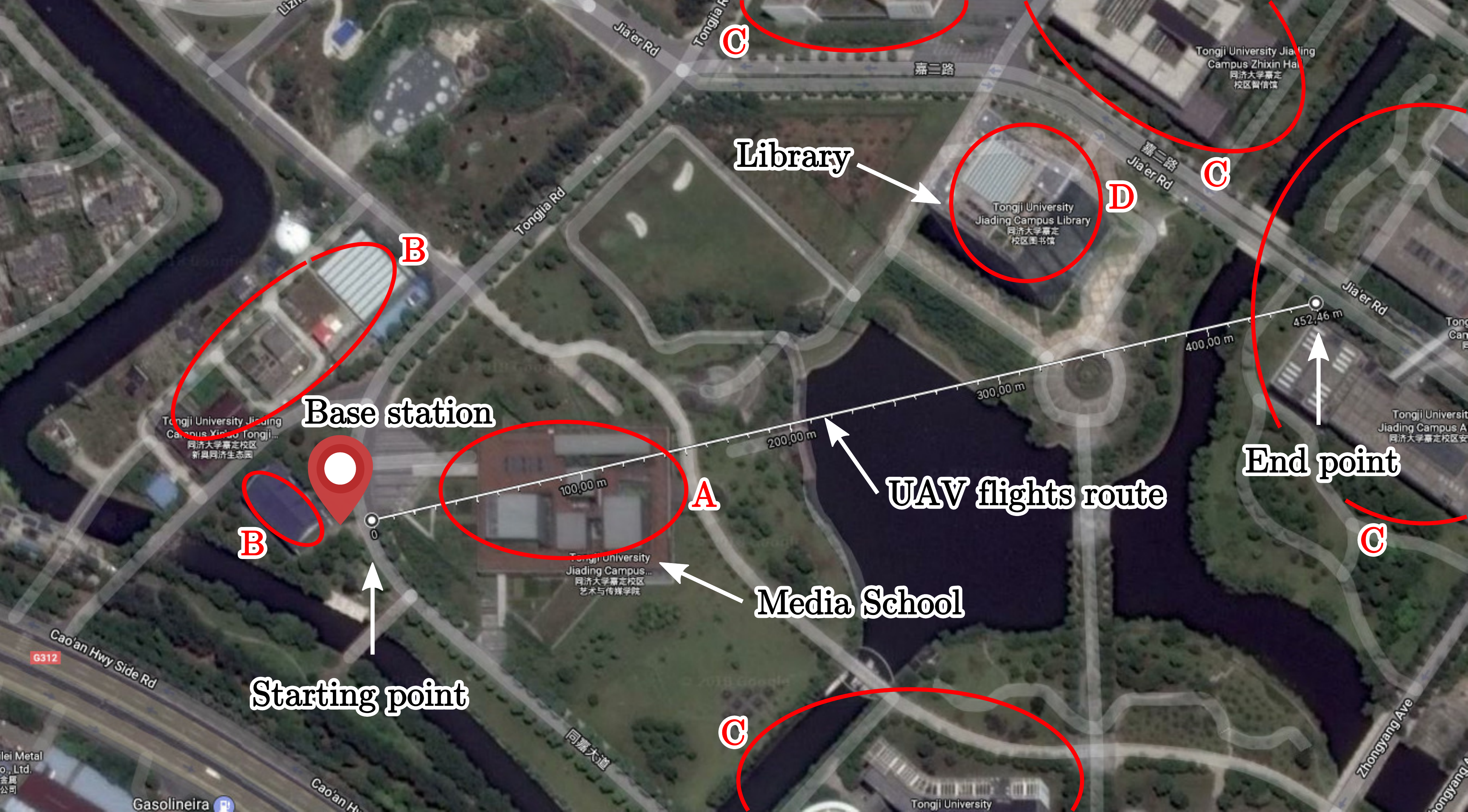}
		\caption{Measurement Environment~II (obstructed \ac{LoS} for lower flight heights, \ac{LoS} \secondMinorReviewXose{otherwise}). \secondReviewCommentXose{This figure was modified.}}
		\label{fig:nlos}
	\end{subfigure}
	\caption{Considered measurement environments.}
	\label{fig:MeasurementScenarios}
\end{figure}

A measurement campaign consisting in horizontal flights at different heights in several suburban scenarios at the Jiading Campus of the Tongji University (Shanghai, China) was performed. Two environments were considered, referred to as Environment~I and Environment~II, imaged respectively in \cref{fig:los,fig:nlos}. The figures also show the position of the transmitter\footnote{Coordinates of the transmitter expressed as (latitude, longitude) are: $\left(31.2873872^\circ, 121.2040907^\circ\right)$.}, as well the flying routes along with their starting and end points\footnote{Coordinates of the starting point and end point expressed as (latitude, longitude) are:  $\left(31.287433^\circ, 121.204179^\circ\right)$ and $\left(31.284102^\circ, 121.208412^\circ\right)$ for the Environment~I, respectively; and $\left(31.287433^\circ, 121.204179^\circ\right)$ and $\left(31.288310^\circ, 121.208793^\circ\right)$ for the Environment~II route, respectively.}. \secondReviewXose{Moreover, buildings with different heights exist in the environment. To approximately describe the building properties, some clusters of buildings are marked in the figures with red ellipses and labeled with the characters from ``A'' to ``D'' based on the corresponding approximate building heights, according to the \cref{table:buildingHeights}.} The so-called Environment~I consists of a \ac{LoS} straight flight along a road without obstacles, being the total length about $560$\,m. For the Environment~II, a $450$\,m-long straight flight in which the \ac{UAV} crosses between several buildings is considered, resulting in an \minorReviewXose{\ac{OLoS}} environment for the lowest height value considered, whereas it can be still categorized as \ac{LoS} for higher flight heights. More specifically, between the values of $50$\,m and $150$\,m (horizontal distance between the \ac{UAV} and the \ac{BS}), the \ac{UAV} flies on top of the Media School building. The height of the building is irregular, but its maximum height point is around $5$\,m lower than the lowest flight height considered. Between the horizontal distance values $300$\,m and $380$\,m the \ac{UAV} flies close to the library, which is approximately $65$\,m high. Both mentioned buildings are also marked in \cref{fig:los,fig:nlos}.

\begin{table}
	\centering
	\begin{small}
		\begin{tabular} {cc}
			\toprule
			\secondReviewXose{Cluster label} & \secondReviewXose{Range of heights [m]}\\
			\midrule
                        \secondReviewXose{A}&\secondReviewXose{$<15$}\\
\secondReviewXose{B}&\secondReviewXose{$15$ -- $25$}\\
\secondReviewXose{C}&\secondReviewXose{$25$ -- $45$}\\
\secondReviewXose{D}&\secondReviewXose{$>60$}\\
			\bottomrule
		\end{tabular}
	\end{small}
	\caption{\secondReviewXose{Approximate height values for the clusters of buildings in \cref{fig:MeasurementScenarios}.}\label{table:buildingHeights}}
\end{table}

For each measurement environment, four horizontal round-trip flights were considered, having a different flight height in the go and the return trip. \cref{table:heightValues} summarizes the height values considered.

\begin{table}
	\centering
	\begin{small}
		\begin{tabular} {ccc}
			\toprule
			Flight no. & Height (go trip) [m] & Height (return trip) [m]\\
			\midrule
                        1&15&25\\
2&35&45\\
3&60&75\\
4&90&105\\
			\bottomrule
		\end{tabular}
	\end{small}
	\caption{Height values for the flights considered.\label{table:heightValues}}
\end{table}

\subsection{Measurement Equipment}\label{sec:MeasurementEquipmentDetails}

\begin{figure}[t!]
	\centering
	\begin{subfigure}[t]{\linewidth}
		\centering
		\includegraphics[width=\columnwidth]{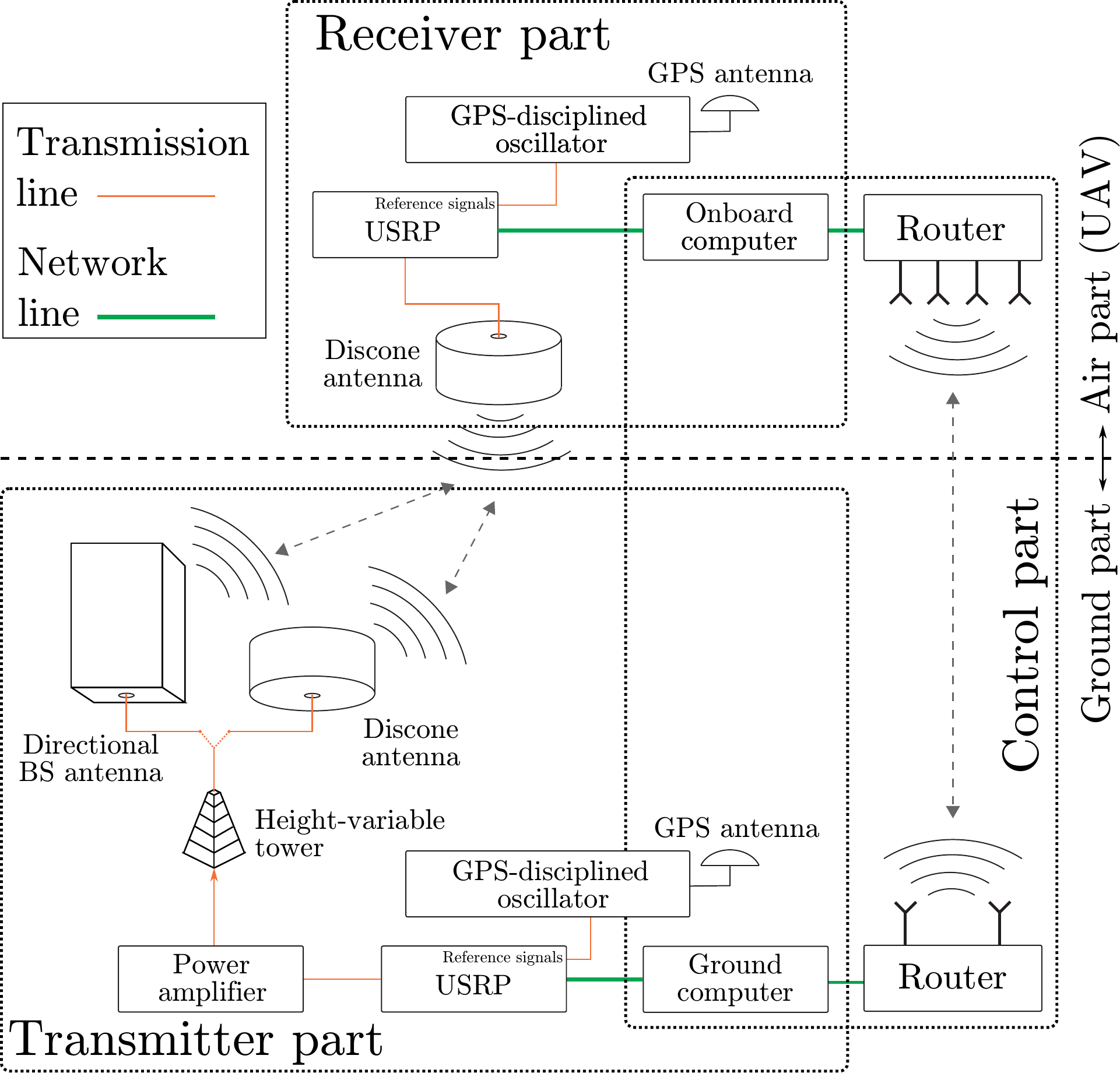}
		\caption{Diagram of the equipment used during the measurements}
		\label{fig:MeasurementEquipment-Diagram}
	\end{subfigure}
	\par\medskip
	\begin{subfigure}[t]{.48\linewidth}
		\centering
		\includegraphics[width=\columnwidth]{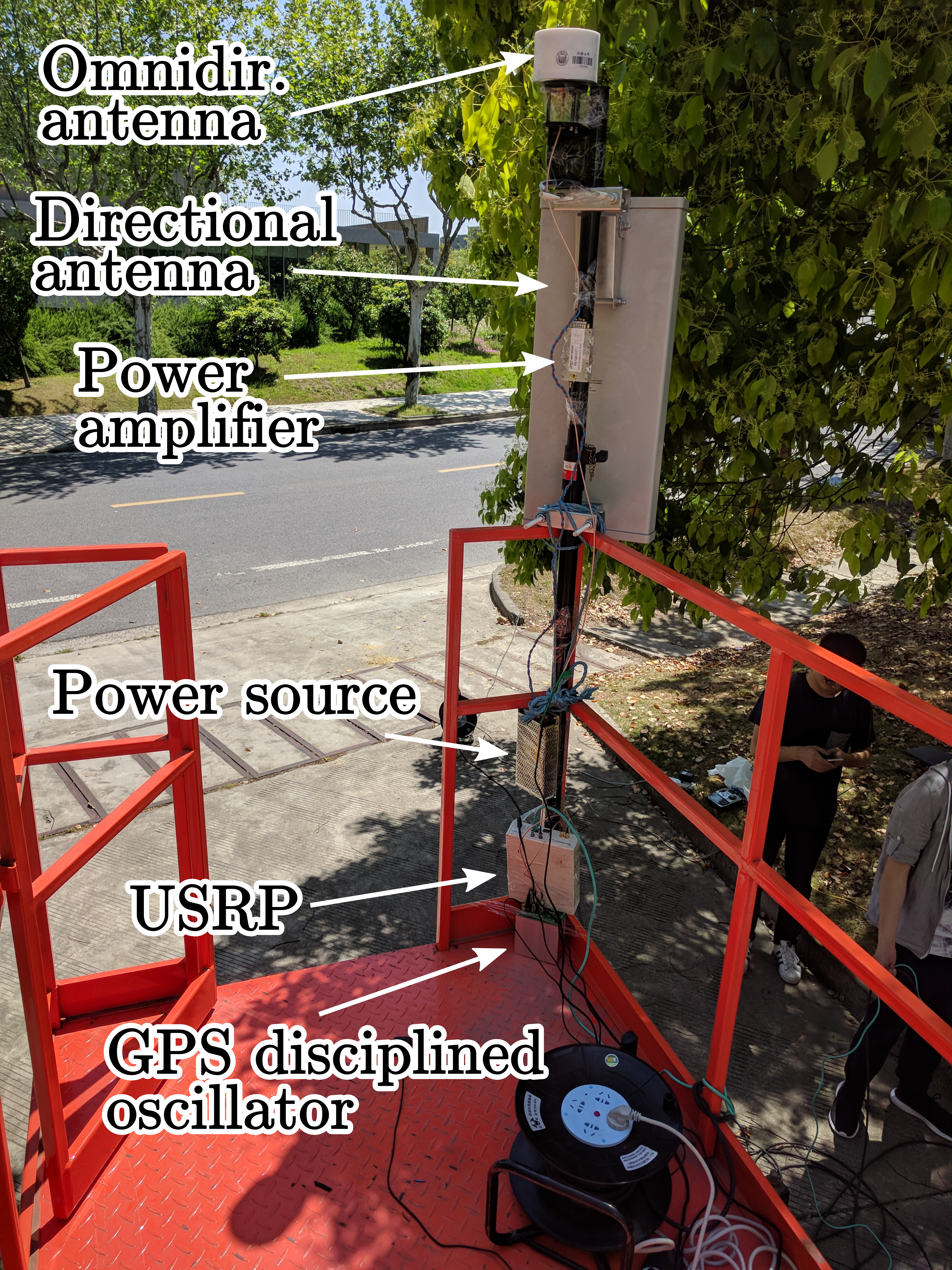}
		\caption{Ground part (transmitter)}
		\label{fig:MeasurementEquipment-TX}
	\end{subfigure}
	\hfill
	\begin{subfigure}[t]{.48\linewidth}
		\centering
		\includegraphics[width=\columnwidth]{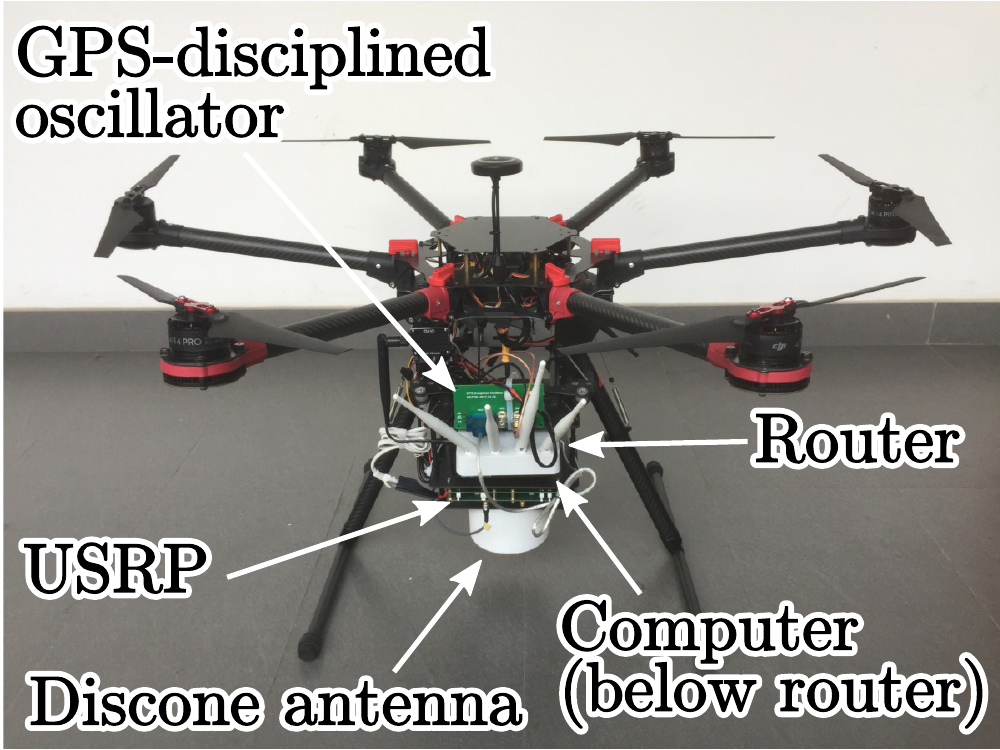}
		\caption[Picture of the air part of the measurement equipment]{Air part (receiver)}
		\label{fig:MeasurementEquipment-RX}
	\end{subfigure}
	\caption{Measurement equipment.}
	\label{fig:Measurement_Equipment_Pictures}
\end{figure}

\begin{figure}[t!]
	\centering
	\begin{subfigure}[t]{.85\linewidth}
		\centering
		\includegraphics[width=\columnwidth]{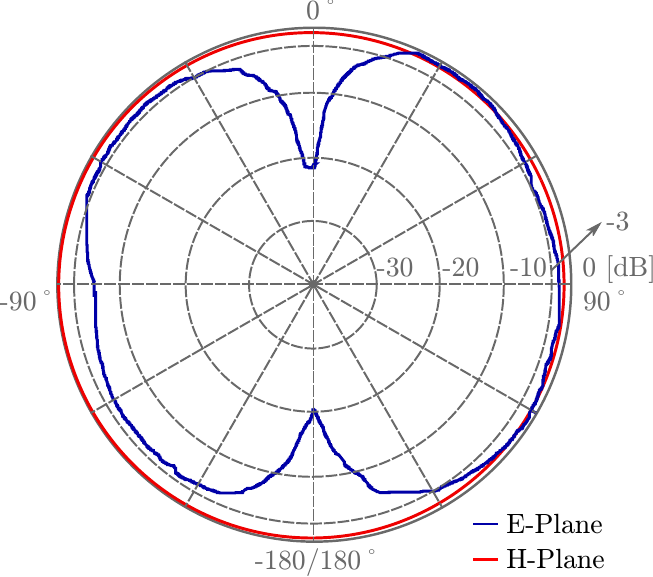}
		\caption{\secondReviewXose{Omnidirectional antenna (used at the \ac{UAV} and tbe \ac{BS}).}}
		\label{fig:RadiationPattern-Omnidirectional}
	\end{subfigure}
	\par\medskip
	\begin{subfigure}[t]{.85\linewidth}
		\centering
		\includegraphics[width=\columnwidth]{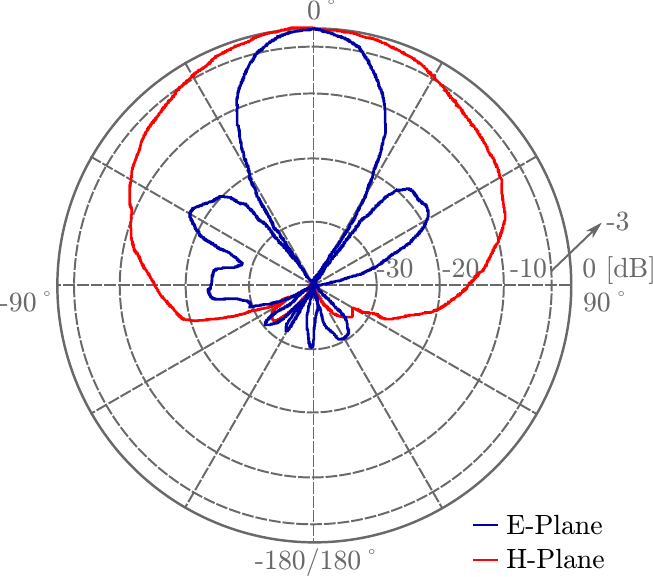}
		\caption{\secondReviewXose{\ac{BS} directional antenna (used at the \ac{BS}).}}
		\label{fig:RadiationPattern-BS}
	\end{subfigure}
	\caption{\secondReviewXose{Radiation patterns of the considered antennas.}}
	\label{fig:RadiationPatterns}
\end{figure}

\cref{fig:MeasurementEquipment-Diagram} illustrates a diagram of the equipment transmitting and acquiring the signals. It consists of two parts, the air part and the ground part. The air part was loaded on a \ac{UAV} as illustrated in \cref{fig:MeasurementEquipment-RX}. It contains the following components: a quasi-omnidirectional packaged discone antenna \secondReviewXose{(see the radiation pattern in \cref{fig:RadiationPattern-Omnidirectional})}, a \ac{USRP} N-210 used to acquire the signals, a \ac{GPS}-disciplined oscillator that generates accurate $10$\,MHz  as well as $1$\,\ac{PPS} reference signals to be provided to the \ac{USRP} for synchronization purposes, a small computer base unit that controls the \ac{USRP} device and stores the received data, and a commercial Wi-Fi router. The ground part contains another \ac{USRP} N-210 used to transmit the signals, a \ac{GPS}-disciplined oscillator, a power amplifier and two transmitter antennas, being only one used at a time (see \cref{fig:MeasurementEquipment-TX}). \reviewXose{Note that the $10$\,MHz and $1$\,\ac{PPS} signals at both ends of the channel (transmitter and receiver) enable synchronization of the used \acp{USRP}. On the one hand, the $10$\,MHz is used as a reference for the internal oscillators of the \acp{USRP}, hence both transmitter and receiver are synchronized in frequency up to the limits of the hardware used, which is essential for the accuracy of the obtained results. On the other hand, $1$\,\ac{PPS} signal enables synchronization of the sample times for the transmitter and the receiver. This way, the transmitter and receiver share a common absolute time-basis. By knowing the time instants in which an \ac{OFDM} frame transmission starts and when it is received, the absolute propagation delay can be estimated.} With the aim of helping the research on the feasibility of serving aerial vehicles using \ac{LTE} network deployments with \ac{BS} antennas targeting terrestrial coverage, two antennas were considered at the transmitter, being one of them the same model mounted at the receiver and the other a typical \ac{BS} directional antenna (see \cref{fig:MeasurementEquipment-TX})\secondReviewXose{, with their respective radiation patterns shown in \cref{fig:RadiationPattern-Omnidirectional,fig:RadiationPattern-BS}}. While the results for the omnidirectional antenna allow for obtaining an accurate characterization of the channel characteristics, results for the directional \ac{BS} antenna show the behavior for \reviewXose{an individual sector of a \ac{BS} (note that, in practical deployments, several sectors per \ac{BS} are set up)}. Both antennas are mounted at a height-variable tower fixed at a height of $15$\,m. Finally, a laptop is connected to the ground \ac{USRP} device to act as a transmitter. Furthermore,  by using another commercial Wi-Fi router, a local area network is established with the computer on the \ac{UAV} allowing to control the on-board equipment remotely. \reviewXose{This connection is only required to be available when the receiver is at the vicinity of the human operator (close to the transmitter) in order to be able to send respective commands to start and stop the signal acquisition, i.e., no permanent connection is required during the whole flight. The receiver is able to work and acquire samples autonomously after it receives a command to start the signal acquisition until it receives a command to stop the acquisition. Note also that the \ac{UAV} flies automatically on preprogrammed routes and hence does not need permanent connection to ground. However, a manual operator followed the \ac{UAV} on ground during the measurements to keep the connection for safety purposes.} It is worth noting that the routers worked at the frequency band of $2.4$\,GHz causing no interference to the measurements. More specifically, the measurements were performed considering a central carrier frequency of $2.5$\,GHz and a bandwidth of $15.36$\,MHz (see \cref{sec:SignalGenerationAndProcessing})\reviewXose{, values which are similar to those corresponding to commercial \ac{LTE} deployments in the area of the measurements\footnote{\reviewXose{Note also that the $2.5$\,GHz band is planned to be used in sub-$6$\,GHz \ac{5G} network deployments in China and other countries.}}}. Moreover, the measurements are geo-localized based on the \ac{GPS} data and the effect of the radiation pattern of the omnidirectional antennas at the \ac{UAV} and the \ac{BS} was compensated. However, the effect of the radiation pattern of the directional antenna used at the \ac{BS} was not removed because we are interested in comparing the results when a directional antenna is considered with those in which the channel is not affected by the antennas (\minorReviewXose{e.g., when an omndirectional antenna is used and its radiation patter\secondMinorReviewXose{n} is compensated}). The directional antenna is oriented so that \minorReviewXose{the axis of} the main lobe follows the direction of the $15$\,m height flight for each scenario. This way, a $0^{\circ}$ tilt configuration was considered, being the main lobe parallel to the ground. \secondReviewXose{\cref{table:radioParameters} details the main parameters of the configuration of the radio equipment used for the measurements.}

\begin{table}
	\centering
	\begin{small}
		\begin{tabular} {cc}
			\toprule
			\secondReviewXose{Parameter} & \secondReviewXose{Value}\\
			\midrule
                        \secondReviewXose{Transmit power}&\secondReviewXose{$40$\,dBm}\\[0.25em]
\multirow{2}{*}{\secondReviewXose{Antenna gain}}&\secondReviewXose{$0$\,dBi (omnidirectional, \ac{UAV} and \ac{BS})}\\
&\secondReviewXose{$12$\,dBi (directional, \ac{BS})}\\[0.25em]
\secondReviewXose{\ac{BS} antenna height}&\secondReviewXose{$15$\,m}\\
			\bottomrule
		\end{tabular}
	\end{small}
	\caption{\secondReviewXose{Configuration parameters of the radio equipment.}\label{table:radioParameters}}
\end{table}


\section{Signal Generation and Processing}\label{sec:SignalGenerationAndProcessing}

Both for the generation as well as processing of the signals, the so-called ``GTEC 5G Simulator'' was used \reviewXose{\cite{ArtigoSimuladorGTEC_IWSLS2016,GTEC_TESTBED_CODE}}. The ``GTEC 5G Simulator'' is a versatile piece of software that enables to fully configure the transmit signal and includes all the necessary developments for processing the acquired samples, such as channel estimation, interpolation and equalization algorithms, as well as time and frequency synchronization\footnote{The source code of both the GTEC Testbed and the GTEC 5G Simulator is publicly available under the GPLv3 license at \url{https://bitbucket.org/tomas_bolano/gtec_testbed_public.git}.}. Recently, a \ac{HRPE} algorithm, namely \ac{SAGE} algorithm, similar to that proposed in \cite{fleury1999channel}, was integrated in the ``GTEC 5G Simulator'' receiver, which allows us to estimate the different parameters of the impinging waves for the acquired signals. More specifically, we consider the delay, the complex-valued amplitude, and the Doppler frequency for each path. For this study, an \ac{OFDM} signal featuring a sampling rate of $15.36$\,MHz was considered. The frame structure is very similar to the one defined for the $10$\,MHz downlink profile of \ac{LTE} \cite{ts_136213p}. Therefore, the estimated time-varying channel impulse response for the $i$th frame is expressed as
\begin{equation}
\label{eq:sage_channel}
h_i(t,\tau) = \sum_{l=1}^{M} \alpha_{i,l} \exp\{j2\pi\nu_{i,l}t\}\delta(\tau-\tau_{i,l}),
\end{equation}
where $t$ is the time variable, $\tau$ is the delay variable, $M$ is the number of waves or paths considered, $\alpha_{i,l} \in \mathbb{C}$ is the $l$th-path amplitude, $\nu_{i,l} \in \mathbb{R}$ and $\tau_{i,l} \in \mathbb{R}$ are the respective Doppler frequency and delay for the $l$th path, and $\delta(\cdot)$ is the Dirac delta function.
We also consider that for $i \neq j$, the $l$th \ac{MPC} is not necessarily the same, i.e., the situation $\nu_{i,l} \neq \nu_{j,l}$ and $\tau_{i,l} \neq \tau_{j,l}$ may be observed. For our study, we estimated $M=15$ paths, which is a number large enough to capture all the \acp{MPC} of the signal in the considered environments. \secondReviewXose{\cref{table:signalParameters} details the main parameters of the signal generation and processing chains considered.}

\begin{table}
	\centering
	\begin{small}
		\begin{tabular} {cc}
			\toprule
			\secondReviewXose{Parameter} & \secondReviewXose{Value}\\
			\midrule
                        \secondReviewXose{Carrier frequency}&\secondReviewXose{$2.5$\,GHz}\\[0.25em]
\multirow{2}{*}{\secondReviewXose{Bandwidth}}&\secondReviewXose{$15.36$\,MHz}\\
&\secondReviewXose{($9$\,MHz without guard band)}\\[0.25em]
\multirow{2}{*}{\secondReviewXose{Sampling frequency}} & \secondReviewXose{$15.36$\,MSamples/s (upsampled to}\\
&\secondReviewXose{$25$\,MSamples/s when transmitting)}\\[0.25em]
\secondReviewXose{FFT size} & \secondReviewXose{$1024$\,points}\\
\secondReviewXose{Used subcarriers} & \secondReviewXose{$600$ (excluding DC)}\\
\secondReviewXose{Subcarrier spacing} & \secondReviewXose{$15$\,kHz}\\
\secondReviewXose{Cyclic prefix length} & \secondReviewXose{$72$\,samples}\\
\secondReviewXose{\secondReviewXose{Estimated paths}} & \secondReviewXose{$15$}\\
\secondReviewXose{\secondReviewXose{Delay resolution}} & \secondReviewXose{$65.1$\,ns}\\
			\bottomrule
		\end{tabular}
	\end{small}
	\caption{\secondReviewXose{Parameters of the signal generation and processing chains.}\label{table:signalParameters}}
\end{table}

\section{Channel Characterization for the Flights}\label{sec:ChannelCharacterizationForTheFlights}

This section shows the results of the statistical channel characterization for both measurement environments and both transmitter antennas considered. For each of the channel characteristics, along with its definition, a sample plot of the obtained results is shown. Unless otherwise specified, the so-called ``sample case'' corresponds to the omnidirectional antenna and the Environment~I, considering all the height values. The channel characteristics are plotted versus the horizontal distance between the \ac{UAV} and the \ac{BS}. We provide both a cloud of points with the individual estimated values per frame (see \cref{eq:sage_channel}) as well a smoothed curve for the sake of clarity. Finally, only the range of horizontal distances in which the speed of the \ac{UAV} is stable (approximately from $100$\,m to $500$\,m) was considered. After graphically showing the results for the sample case, statistical fittings are proposed for all the combinations of flight heights, measurement environment and transmitter antenna; and a discussion on the obtained results is provided. Note that in some cases, the results for different height values can be well described by a common statistical distribution. This way, in some cases we consider a single statistical distribution for certain flight height range.  Furthermore, for the sake of clearness, only some sample empirical \acp{CDF}, as well as their fittings, are shown in the paper.

\cref{sec:PathLoss,sec:ShadowFading} consider the path loss and the shadow fading, respectively, whereas \cref{sec:PdpAndDPSD} considers the \ac{PDP} and Doppler frequency \ac{PSD}. \cref{sec:rmsDelaySpread,sec:rmsDopplerFrequencySpread} consider the \ac{RMS} delay and Doppler frequency spreads, respectively, and \cref{sec:riceanKFactor} considers the Ricean K-factor.

\subsection{Path Loss}\label{sec:PathLoss}

The path loss is the ratio between the transmitted and the received power, given by \cite{rappaport1996wireless} in decibels as
\begin{equation}
\label{eq:pathloss}
\text{PL}(d) = 10\log_{10}\left(\frac{P_t}{P_r(d)}\right),
\end{equation}
where $\text{PL}(d)$ is the path loss for a distance $d$, $P_t$ is the transmit power, and $P_r(d)$ is the received power at a distance $d$. The path loss can be modeled by a simple log-distance model \cite{rappaport1996wireless} as
\begin{equation}
\label{eq:pathloss-model}
\begin{split}
\text{PL}(d) &= \overline{\text{PL}}(d_0) + 10 \gamma\log_{10}\left(\frac{d}{d_0}\right) + X_\sigma\\
&= b + 10 \gamma \log_{10} d + X_\sigma,
\end{split}
\end{equation}
where $d_0$ is the so-called ``break distance'' (a reference distance relatively close to the transmitter \cite{rappaport1996wireless}), $\overline{\text{PL}}(d_0)$ is the mean path loss at the distance $d_0$, $\gamma$ is the path loss exponent, $X_\sigma$ is a zero-mean Gaussian random variable, and $b = \overline{\text{PL}}(d_0) - 10\gamma\log_{10}(d_0)$.

\begin{figure}[t!]
	\centering
	\begin{subfigure}[t]{\linewidth}
		\centering
		\includegraphics[width=\columnwidth]{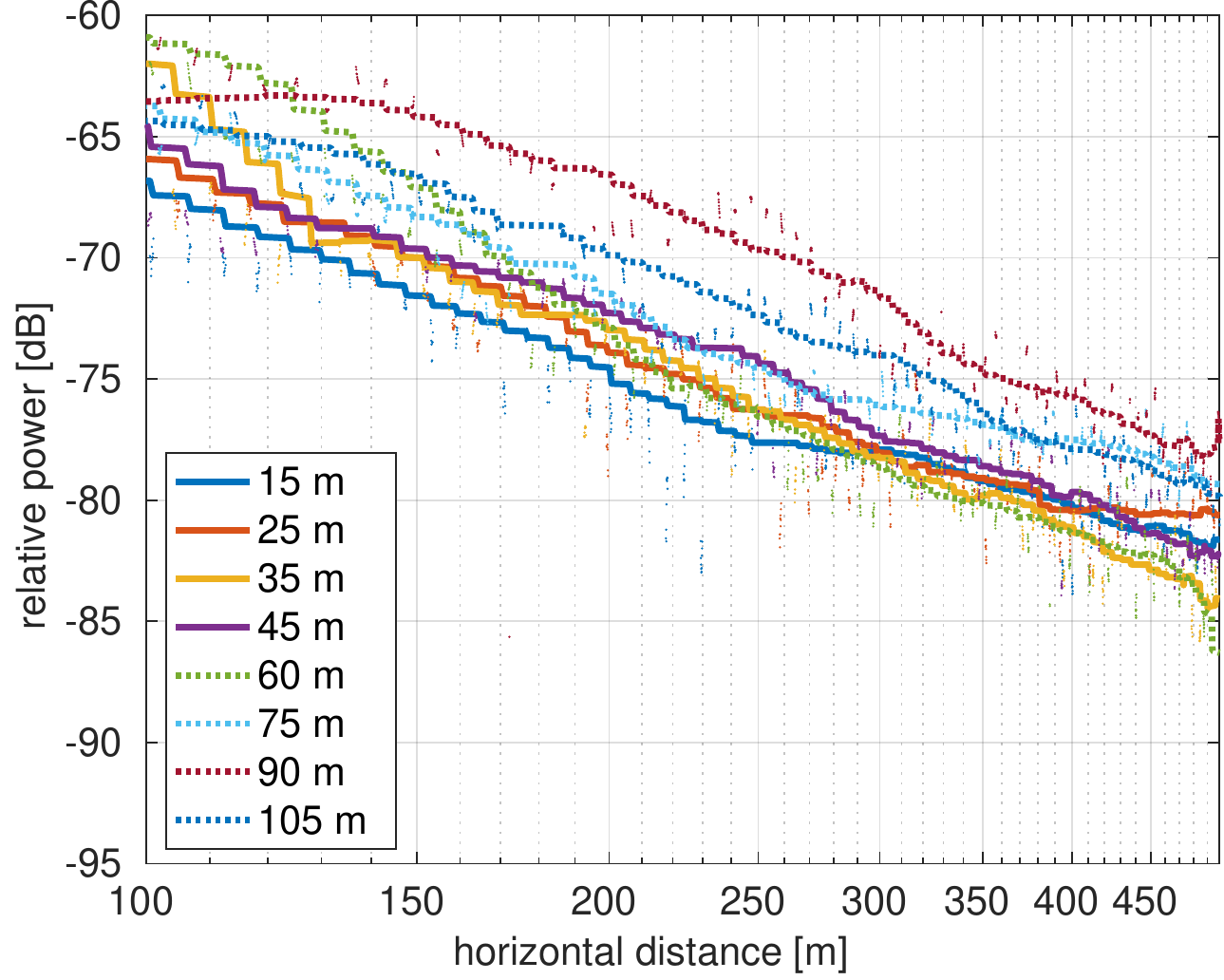}
		\caption{Relative received power.}
		\label{Fig:power-2500MHzLoS-omni}
	\end{subfigure}
	\par\medskip
	\begin{subfigure}[t]{\linewidth}
	\centering
	\includegraphics[width=\columnwidth]{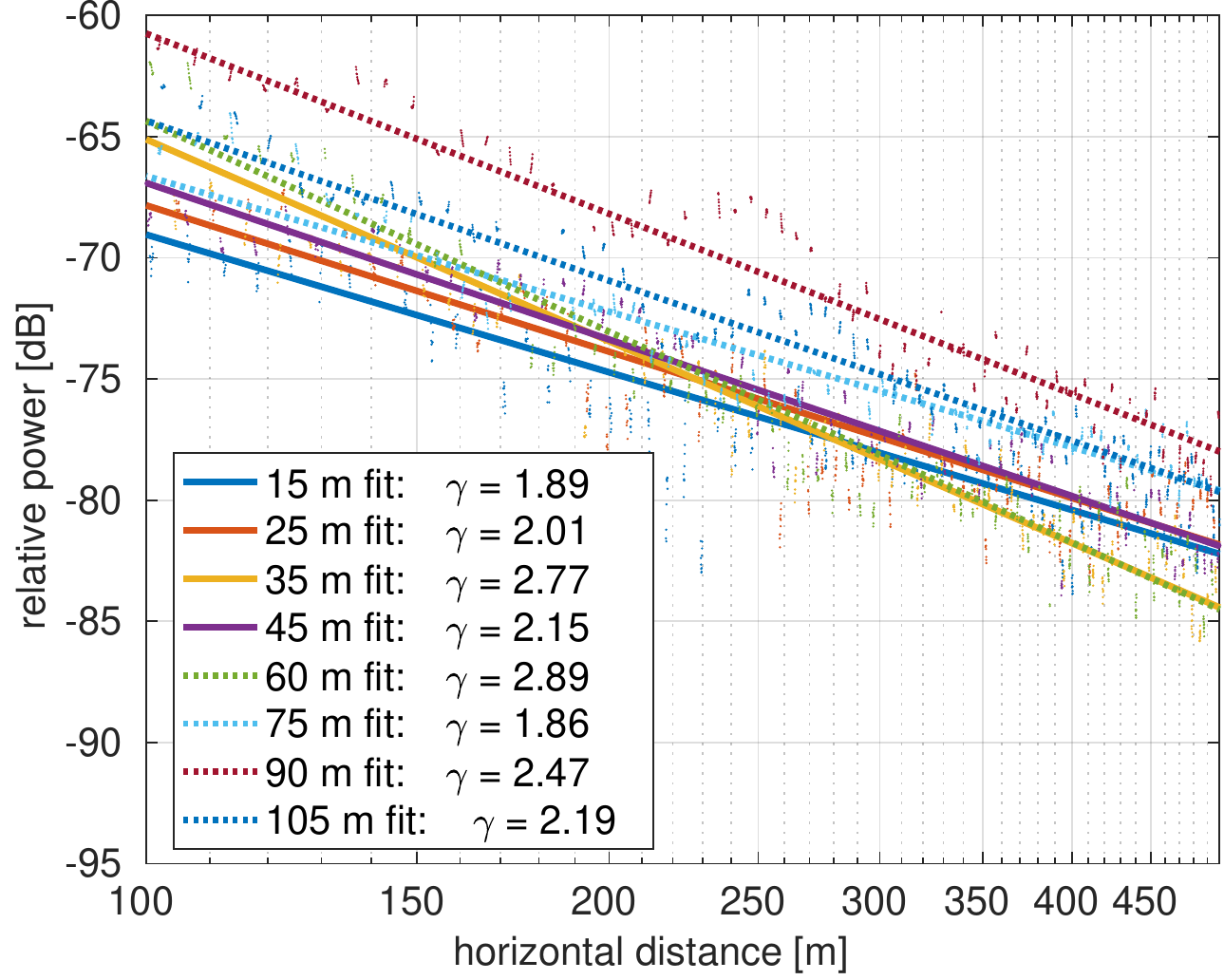}
	\caption{Fittings for the relative received power.}
	\label{Fig:powerfit-2500MHzLoS-omni}
	\end{subfigure}
	\caption{Relative received power values and their fittings obtained at different heights when the omnidirectional antenna is used at the transmitter and the Environment~I is considered.}
	\label{fig:figurePathLossResults}
\end{figure}

\cref{Fig:power-2500MHzLoS-omni} shows the relative received power values for the sample case, whereas \cref{Fig:powerfit-2500MHzLoS-omni} shows the corresponding linear fit for each of the relative received power clouds, obtained by applying a robust linear fit method \cite{Huber81}. All the obtained path loss exponents are detailed in \cref{table:PathLoss}, for both transmitter antennas as well as for the two environments considered.

It can be seen that the path loss exponent results are quite dependent with both the height and the environment considered. The transmit antenna has also a great influence on the results. When the omnidirectional antenna is used, the path loss exponent increases with the flight height reaching a maximum at $60$\,m and then slightly decreases with the height. \reviewXose{For the case of the Environment~II, two exponents are calculated when the flight height is $15$\,m, labeled as \ac{LoS} and \ac{OLoS}, respectively\footnote{\reviewXose{Note that there are no \ac{OLoS} results in the models corresponding to the Environment~I since, for this environment, all the flights are performed in \ac{LoS} conditions.}}. This is due to the effect of partial blockage of the low-height building overflight by the \ac{UAV} (see \cref{fig:nlos}), which intrudes the Fresnel Area. This way, a sharp drop in the receiver power for both the omnidirectional and directional transmit antennas is exhibited after the \ac{UAV} crosses the building (at around a horizontal distance of $170$\,m from the \ac{BS}), impacting severely on the path loss exponent. Hence, the \ac{LoS} path loss exponent corresponds to the propagation before crossing the building and the \ac{OLoS} for the rest of the \ac{UAV} flight route.} For the cases when the directional antenna is used, the results are less regular. In general, it can be seen that the path loss exponent is larger for moderate heights (from $15$\,m to \reviewXose{$35$\,m}), and severely decreases for higher flights  \reviewXose{(those whose heights exceed $35$\,m), obtaining values lower than $2.00$ (marked with a star in \cref{table:PathLoss}). This is due to the fact that, for moderate values of horizontal distance, the \ac{UAV} does not fall within the main lobe of the radiation pattern of the directional antenna but still receives eventual contributions from the second lobe (see \cref{fig:RadiationPattern-BS}). When the horizontal distance increases, the \ac{UAV} can receive contributions from the main lobe of the antenna. Hence, the relative received power is not monotonically decreasing with the horizontal distance to the \ac{BS} and the log-distance model for the path loss cannot track the instantaneous effect of the combined channel plus antenna radiation pattern for the whole flight. In this case, a model that includes the effect of the \ac{BS} antenna radiation pattern would be required. However, it must be considered that the absolute received power for these flights is quite low, indicating that a deployment based on terrestrial antennas would need multiple sectors with directional antennas featuring different orientations to be able to cover a large range of flight heights in practice.}


\reviewXoseRemoveText{Taking into account the configuration of the directional antenna at the BS (see Section II-A), for moderate height values, the UAV is expected to fall within the main lobe of the radiation pattern. However, for larger heights, the path loss exponent is reduced but also the average received power is greatly reduced. This is expectable since the UAV is not covered by the main beam of the BS antenna.} 

\begin{table}
	\centering
	\begin{small}
		\begin{tabular} {ccccc}
			\toprule
			Height & Env.~I & Env.~I & Env.~II & Env.~II\\
			{}[m] & Omnidir. ant. & Dir. ant & Omnidir. ant. & Dir. ant\\
			\midrule
                        \reviewXose{$15$ (LoS)} & \reviewXose{$ 1.89$}& \reviewXose{$ 2.36$}& \reviewXose{$ 2.06$}& \reviewXose{$ 2.03$}\\
\reviewXose{$15$ (OLoS)} & \reviewXose{--}& \reviewXose{--}& \reviewXose{$ 3.07$}& \reviewXose{$ 3.73$}\\
$25$ & $ 2.01$& $ 2.12$& $ 1.24$& $ 3.02$\\
$35$ & $ 2.77$& $ 1.98$& $ 1.91$& $ 3.76$\\
$45$ & $ 2.15$& \reviewXose{$ 1.61^{\left(*\right)}$}& $ 2.15$& \reviewXose{$ 1.56^{\left(*\right)}$}\\
$60$ & $ 2.89$& \reviewXose{$ 0.91^{\left(*\right)}$}& $ 3.00$& \reviewXose{$ 0.31^{\left(*\right)}$}\\
$75$ & $ 1.86$& \reviewXose{$ 1.01^{\left(*\right)}$}& $ 2.35$& \reviewXose{$ 0.53^{\left(*\right)}$}\\
$90$ & $ 2.47$& \reviewXose{$ 0.88^{\left(*\right)}$}& $ 1.98$& \reviewXose{$ 0.59^{\left(*\right)}$}\\
$105$ & $ 2.19$& \reviewXose{$ 0.71^{\left(*\right)}$}& $ 2.00$& \reviewXose{$ 1.91^{\left(*\right)}$}\\
			\bottomrule
		\end{tabular}
	\end{small}
	\caption{Path loss exponents obtained for the different environments and transmit antennas. \reviewXose{$^{(*)}$Due to the effect of the radiation pattern of the \ac{BS} antenna, the relative received power is not monotonically decreasing with the horizontal distance and the log-distance model for the path loss cannot track the instantaneous effect of the combined channel plus antenna radiation pattern for the whole flight.}\label{table:PathLoss}}
\end{table}

\subsection{Shadow Fading}\label{sec:ShadowFading}

\begin{figure}[t!]
	\centering
	\begin{subfigure}[t]{\linewidth}
		\centering
		\includegraphics[width=\columnwidth]{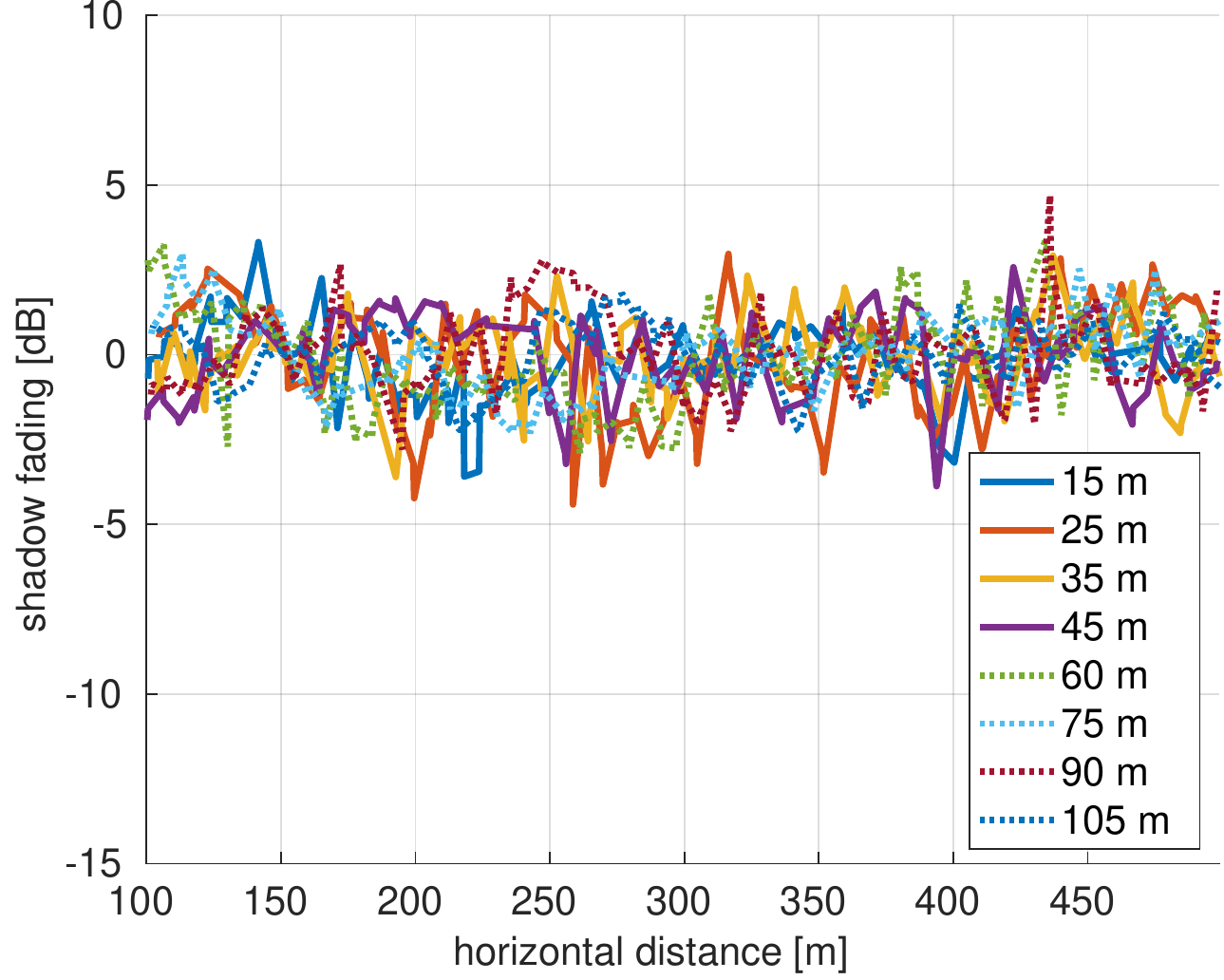}
		\caption{Shadow fading.}
		\label{Fig:sf-2500MHzLoS-omni}
	\end{subfigure}
	\par\medskip
	\begin{subfigure}[t]{\linewidth}
		\centering
		\includegraphics[width=\columnwidth]{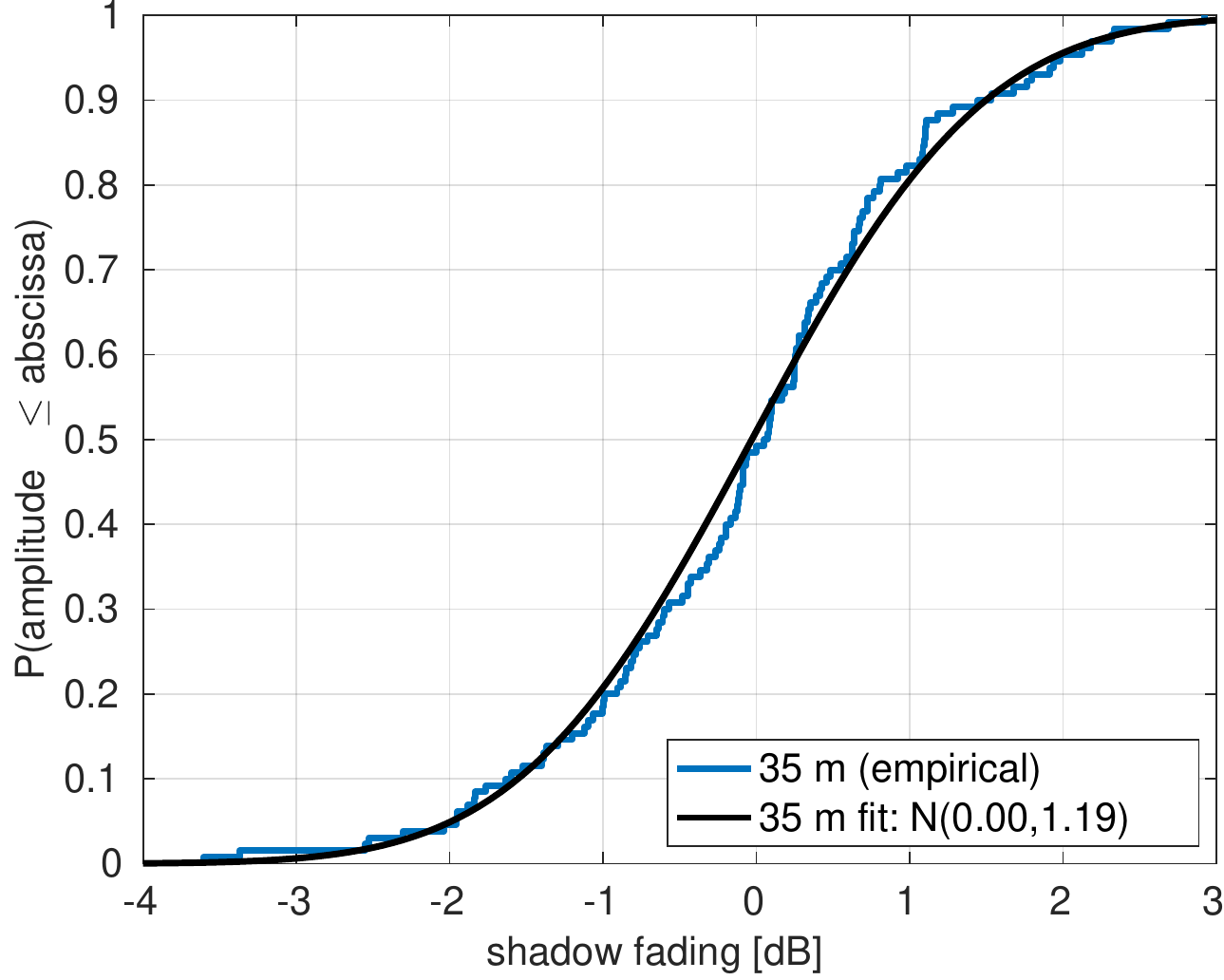}
		\caption{Fitted and empirical CDFs for the shadow fading at the flight height $15$\,m.}
		\label{Fig:cfsf-2500MHzLoS-omni}
	\end{subfigure}
	\caption{Shadow fading results and their fittings obtained at different heights when the omnidirectional antenna is used at the transmitter and the Environment~I is considered.}
	\label{fig:figureShadowFadingResults}
\end{figure}


The shadow fading is calculated by subtracting the path loss from the smoothed received power.  \cref{Fig:sf-2500MHzLoS-omni} shows the shadow fading for the sample case. Not clear dependency with the distance to the \ac{BS} is appreciated in the shadow fading results. \cref{Fig:cfsf-2500MHzLoS-omni} shows the corresponding empirical \ac{CDF} as well as its fitting when the height is $35\text{\,m}$. In general, all the shadow fading values can be fit by a normal distribution with zero mean and the variance values specified in \cref{table:PathShadowFading}. It can be seen that for the cases in which the omnidirectional antenna is used at the \ac{BS} the shadow fading tends to decrease with the flight height, since the channel becomes more \ac{LoS}-alike. For the lowest height values, the shadow fading is \minorReviewXose{slightly} increased due to the effects of the ground elements. This effect is also appreciated for the case of the directional antenna. However, then the directional antenna is used, the shadow fading is not decreased with the flight height. This is caused by the influence of the sidelobes of the \ac{BS} antenna. \reviewXose{Finally, an increase in the shadow fading standard deviation can be appreciated when the \ac{UAV} crosses the media building, hence changing from \ac{LoS} propagation conditions to \ac{OLoS}. This is due to the richer scattering of the \ac{OLoS} scenarios with respect to the \ac{LoS} ones.}

\begin{table}
	\centering
	\begin{small}
		\begin{tabular} {cccccc}
			\toprule
			Height&Env.~I & Env.~I & Env.~II & Env.~II\\
			{}[m]&Omnidir. ant. & Dir. ant & Omnidir. ant. & Dir. ant\\
			\midrule
                        \reviewXose{$15$ (LoS)} & $ 1.35$& $ 2.45$& \reviewXose{$ 1.28$}& \reviewXose{$ 2.77$}\\
\reviewXose{$15$ (OLoS)} & \reviewXose{--}& \reviewXose{--}& $ \reviewXose{1.45}$& \reviewXose{$ 3.04$}\\
$25$ & $ 1.59$& $ 2.66$& $ 1.32$& $ 1.63$\\
$35$ & $ 1.19$& $ 3.01$& $ 1.34$& $ 2.57$\\
$45$ & $ 1.25$& $ 2.50$& $ 1.01$& $ 1.73$\\
$60$ & $ 1.24$& $ 3.01$& $ 1.27$& $ 2.61$\\
$75$ & $ 1.16$& $ 2.47$& $ 1.02$& $ 1.95$\\
$90$ & $ 1.06$& $ 2.53$& $ 1.05$& $ 2.39$\\
$105$ & $ 0.91$& $ 3.06$& $ 0.83$& $ 2.40$\\
			\bottomrule
		\end{tabular}
	\end{small}
	\caption{\minorReviewXose{Standard deviations (in dB)} for the shadow fading fittings in different environments and for different transmit antennas.\label{table:PathShadowFading}}
\end{table}

\subsection{Power Delay Profile and Power Spectral Density}\label{sec:PdpAndDPSD}

The \ac{PDP} contains information about how much power arrives at the receiver with a certain delay $\tau$. In practice, the \ac{PDP} is obtained as the power for a certain timespan over which the channel is quasi-stationary \cite{molisch2012wireless}. Following the same approach as in \cite{ArtigoMedidasMetro_Measurements2017}, we calculate the \ac{PDP} for each acquired frame from the channel estimates produced by the \ac{SAGE} algorithm (see \cref{eq:sage_channel}), thus obtaining the ``instantaneous'' \ac{PDP}. Following \cite{meijerink2014physical}, we define the ``instantaneous'' PDP for the $i$th frame as
\begin{equation}
\label{eq:pdp-sage}
P_i(\tau)  = \sum_{l=1}^M \left|\alpha_{i,l}\right|^2 \delta\left(\tau-\tau_{i,l}\right).
\end{equation}

\begin{figure}[t!]
	\centering
	\begin{subfigure}[t]{\linewidth}
		\centering
		\includegraphics[width=\columnwidth]{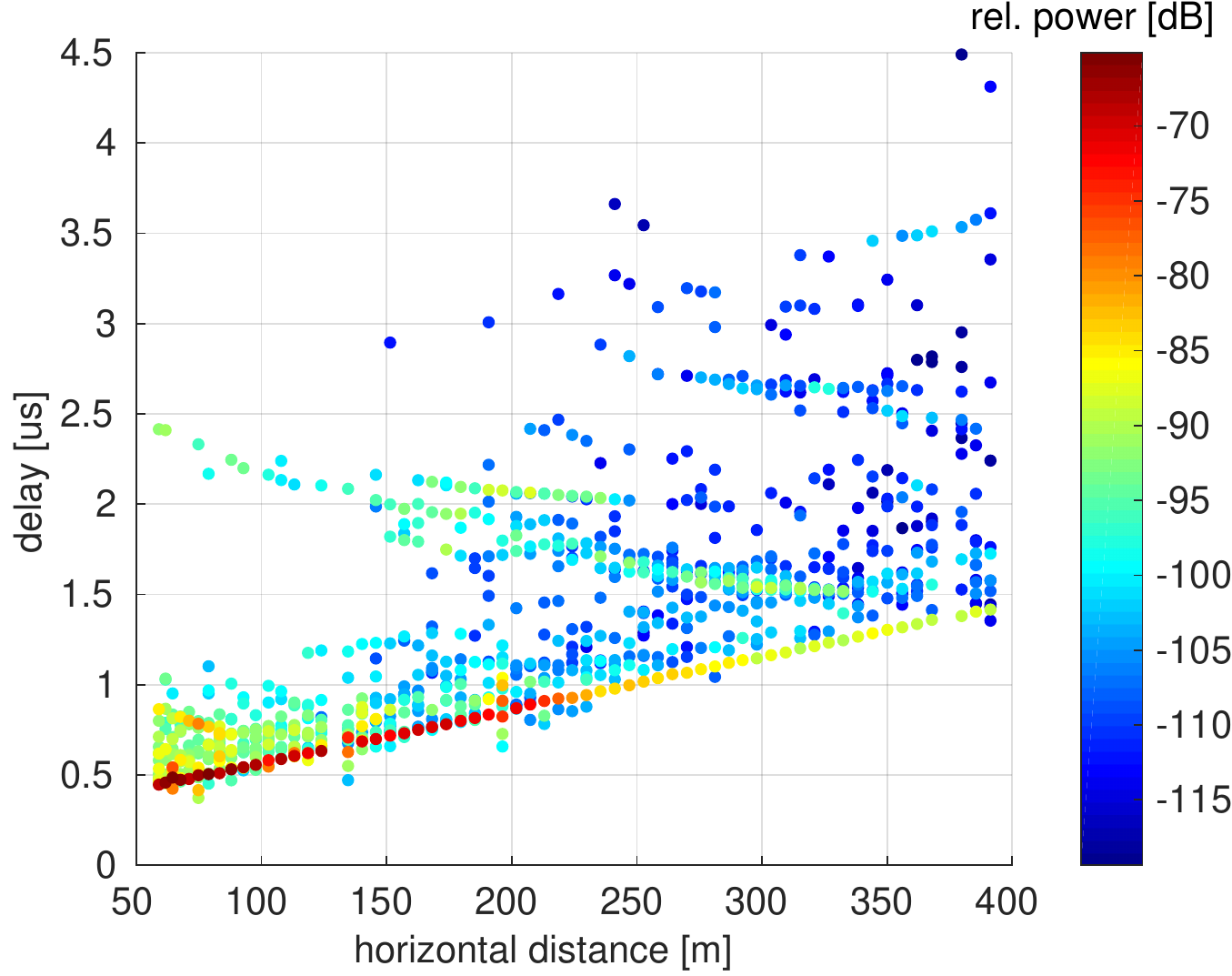}
		\caption{Power Delay Profile. \reviewCommentXose{Units were added to the colorbar.}}
		\label{Fig:pdp-2500MHzNLoS-omni-15}
	\end{subfigure}
	\par\medskip
	\begin{subfigure}[t]{\linewidth}
		\centering
		\includegraphics[width=\columnwidth]{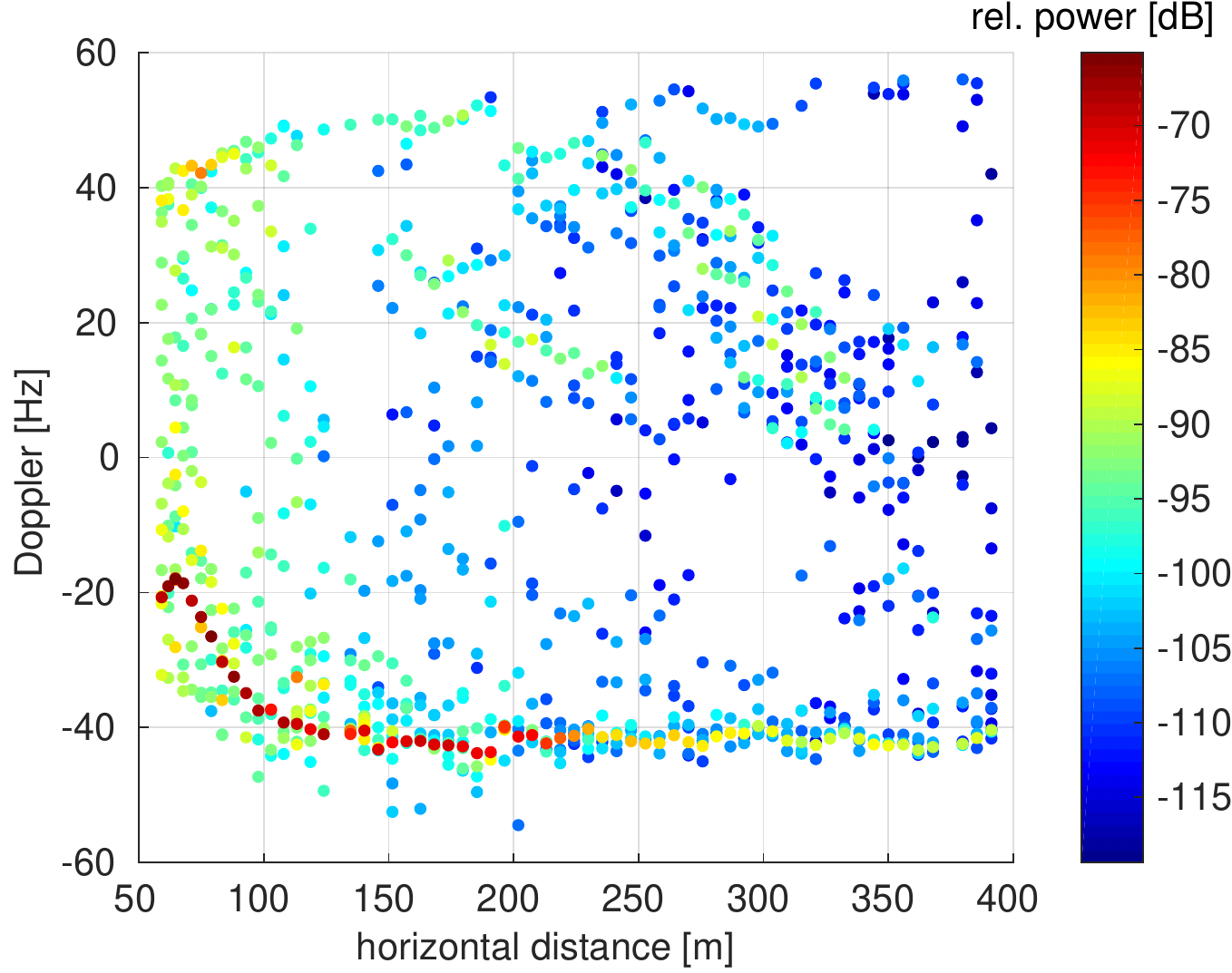}
		\caption{Doppler frequency Power Spectral Density. \reviewCommentXose{Units were added to the colorbar.}}
		\label{Fig:dpsd-2500MHzNLoS-omni-15}
	\end{subfigure}
	\caption{Power Delay Profile and Doppler frequency Power Spectral Density when the omnidirectional antenna is used at the transmitter and the Environment~II is considered, being the flight height $15$\,m.}
	\label{fig:figurePDPeDPSD-Scenario-II}
\end{figure}

As an example, \cref{Fig:pdp-2500MHzNLoS-omni-15} shows the \ac{PDP} when the omnidirectional antenna is used at the transmitter and the Environment~II is considered, for a flight height of $15$\,m, versus the horizontal distance between the \ac{BS} and the \ac{UAV}, for the whole distance range available. This particular flight was considered as an example because it is one of the most rich in scattering components. The points in the figure represent the \acp{MPC} and their color define the relative power values. As expected, the delay values increase with the distance between the \ac{UAV} and the \ac{BS}. It can be seen that, apart from the main component (\ac{LoS}), other well-structured lines of \acp{MPC} can be appreciated, as a result of reflections on elements of the environment. 
Analogously to the \ac{PDP} case, the Doppler \ac{PSD} function contains information about the power of the signals impinging the receiver with a given Doppler frequency. More specifically, the Doppler \ac{PSD} is related to the \acp{AoA} of the \acp{MPC}.
As we did for the \ac{PDP}, we calculate the Doppler \ac{PSD} for each received frame from the channel estimates defined in \cref{eq:sage_channel} and determined with the SAGE algorithm. Following an equivalent approach to the one shown in \cite{meijerink2014physical} for the \ac{PDP}, we define the ``instantaneous'' Doppler \ac{PSD} for the $i$th frame as \cite{ArtigoMedidasMetro_Measurements2017}:
\begin{equation}
\label{eq:doppler-psd-sage}
D_i(\nu) = \sum_{l=1}^{M} \left|\alpha_{i,l}\right|^2 \delta\left(\nu-\nu_{i,l}\right).
\end{equation}

\cref{Fig:dpsd-2500MHzNLoS-omni-15} shows the Doppler frequency \ac{PSD} for the same flight previously considered. The values of the largest components of the Doppler PSD are coherent with the low speed considered for the \ac{UAV} of $5\text{\,m/s}$. While for the go flights (the ones starting from the \ac{BS}) the sign of the dominant Doppler components is negative, the return flights (i.e., those starting from most far away points from the \ac{BS}) exhibit positive values, as expected. The most powerful points in the figure correspond to the contribution of the main (\ac{LoS}) component and their Doppler values accounts for the \ac{UAV} speed (it can be seen that the speed at the beginning of the flight is still increasing until reaching a stable value).


Based on the \ac{PDP} and Doppler frequency \ac{PSD}, we can obtain the \ac{RMS} delay spread and the \ac{RMS}  Doppler frequency spread, respectively.

\subsection{Root Mean Square Delay Spread}\label{sec:rmsDelaySpread}

The \ac{RMS} delay spread, under some circumstances, it is proportional to the error probability due to the delay dispersion \cite{molisch2012wireless}. The delay spread is inversely related with the channel coherence bandwidth, hence a high delay spread will affect the cyclic prefix length and produce \minorReviewXose{\ac{ISI}}, requiring the use of more advanced equalization architectures at the receiver. The \ac{RMS} delay spread is calculated as the normalized second-order central moment of the delay \cite{molisch2012wireless}. Following the same approach as in \cite{ArtigoMedidasMetro_Measurements2017}, let us firstly define the normalized PDP for the $i$th frame as
\begin{equation}
\label{eq:normalized-pdp}
\tilde{P}_i(\tau) = \frac{P_i(\tau)}{\int_{-\infty}^{\infty}P_i(\tau) \dif \tau} =
\frac{\sum_{l=1}^{M}  \left|\alpha_{i,l}\right|^2 \delta\left(\tau-\tau_{i,l}\right)}
{\sum_{l=1}^{M} \left|\alpha_{i,l}\right|^2},
\end{equation}
being $P_i(\tau)$ given by \cref{eq:pdp-sage}. From the result in \cref{eq:normalized-pdp}, the $n$th  moment of the delay is
\begin{equation}
\label{eq:n-moment-pdp}
\mathbb{E}_i[\tau^n] = \int_{-\infty}^{\infty}   \tilde{P}_i(\tau) \tau^n \dif \tau =
\frac{\sum_{l=1}^{M}  \left|\alpha_{i,l}\right|^2 \tau_{i,l}^n}
{\sum_{l=1}^{M} \left|\alpha_{i,l}\right|^2},
\end{equation}
and the \ac{RMS} delay spread is defined as
\begin{equation}
\label{eq:rms-delay-spread}
S_i = \sqrt{\mathbb{E}_i[\tau^2] - \mathbb{E}_i[\tau]^2}.
\end{equation}

\begin{figure}[t!]
	\centering
	\begin{subfigure}[t]{\linewidth}
		\centering
		\includegraphics[width=\columnwidth]{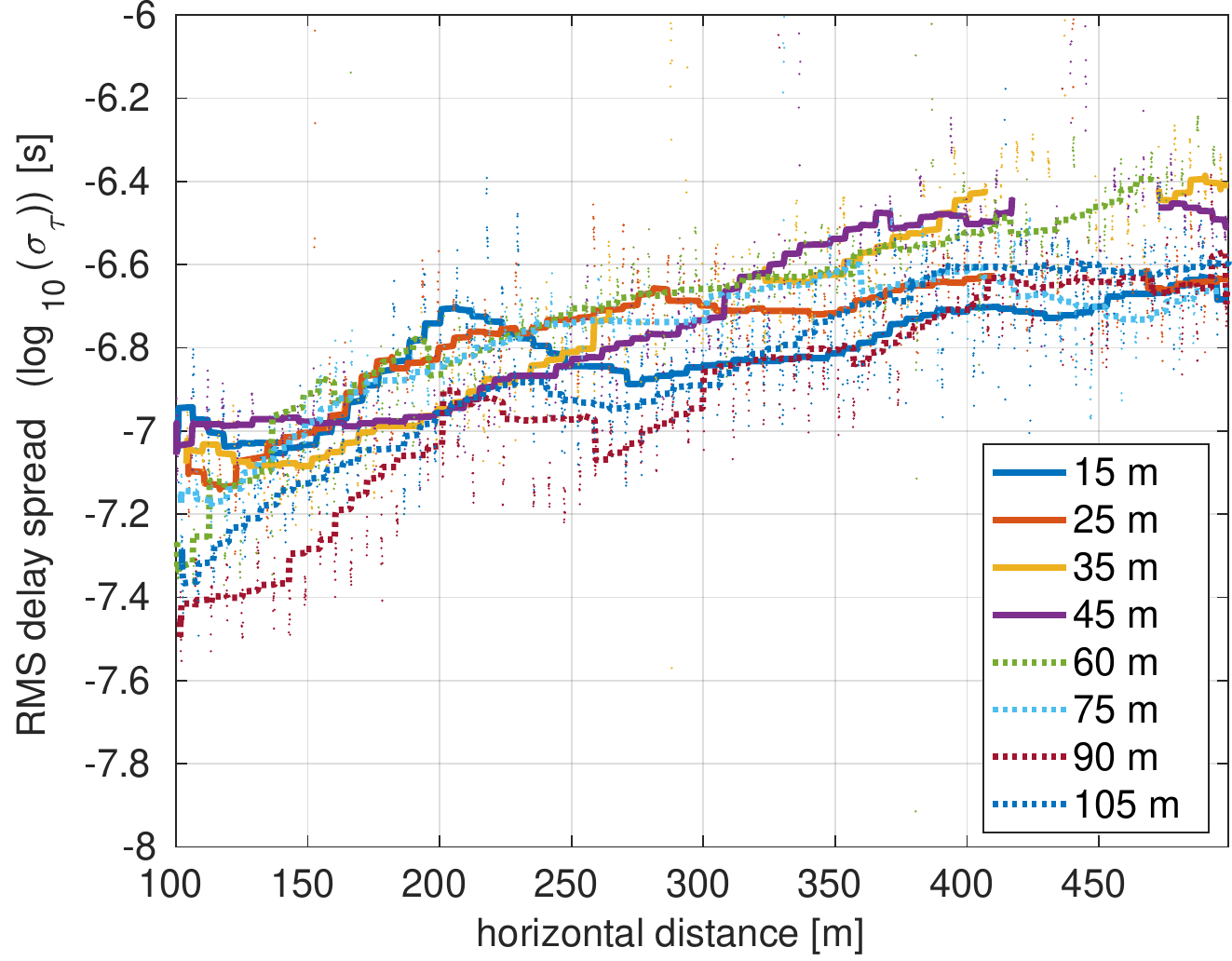}
		\caption{RMS delay spread.}
		\label{Fig:ds-2500MHzLoS-omni}
	\end{subfigure}
	\par\medskip
	\begin{subfigure}[t]{\linewidth}
		\centering
		\includegraphics[width=\columnwidth]{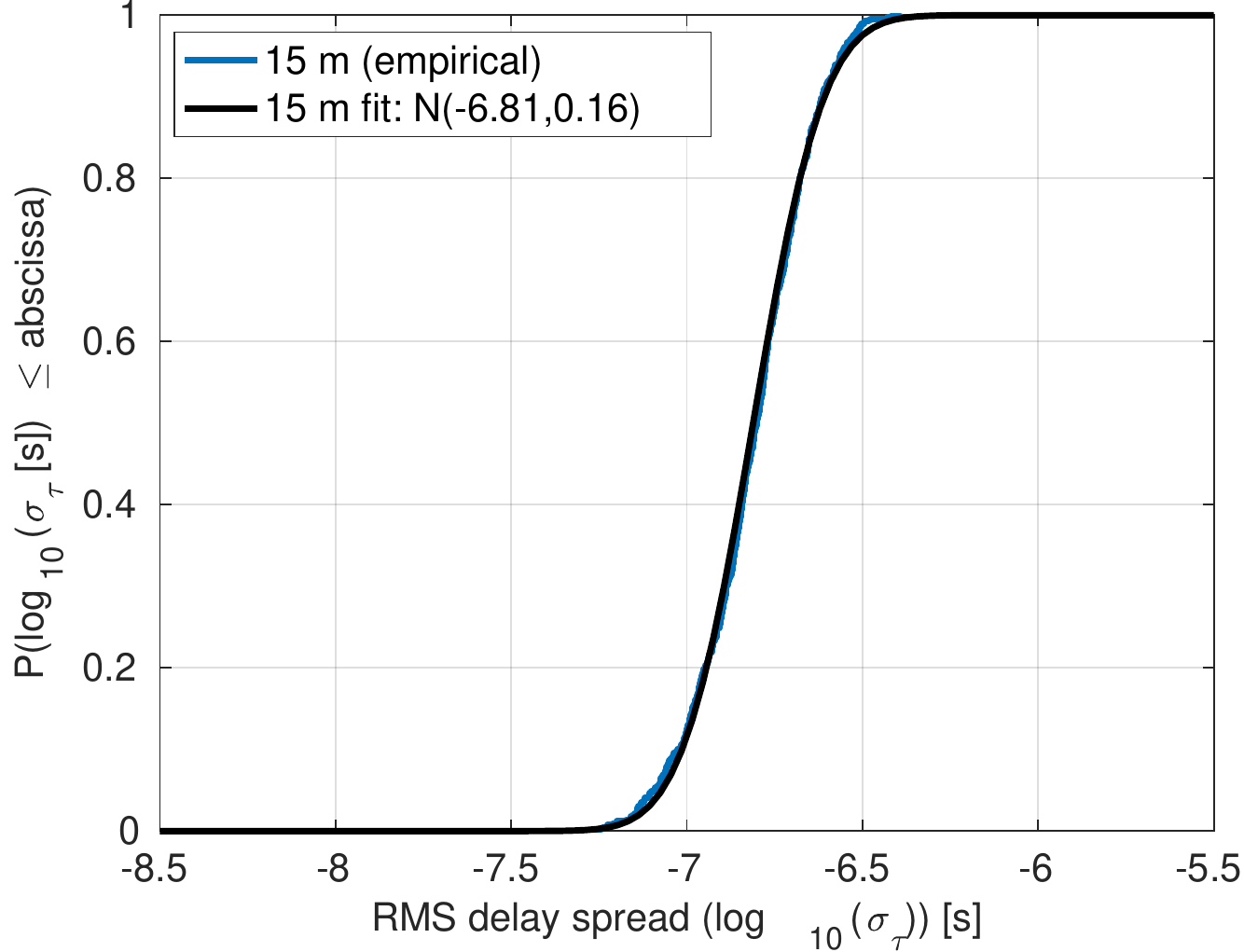}
		\caption{RMS delay spread fitting for the flight height  $15$\,m.}
		\label{Fig:cfds-2500MHzLoS-omni}
	\end{subfigure}
	\caption{RMS delay spread for different heights and sample fittings when the omnidirectional antenna is used at the transmitter and the Environment~I is considered.}
	\label{fig:figureDS}
\end{figure}

\cref{Fig:ds-2500MHzLoS-omni} shows the \ac{RMS} delay spread values for the sample case. There is not a strong dependency with the flight height, although in general the values are slightly lower for the highest heights. It can be seen that, for all the cases, the \ac{RMS} delay spread increases with the distance to the \ac{BS}. This can be explained because the \ac{LoS} \ac{MPC} becomes less powerful when the distance increases (e.g., see \cref{Fig:pdp-2500MHzNLoS-omni-15}). \cref{Fig:cfds-2500MHzLoS-omni} shows the proposed fittings for some of the empirical \acp{CDF} for each cloud of points of the RMS delay spread values. In \cref{table:DelaySpread} we summarize the obtained results for the different environments and transmit antennas. In all cases, normal distributions are used for the fittings. When the directional \ac{BS} antenna is used, the \ac{RMS} delay spread for mid-height flights results slightly lower, since the \ac{LoS} path is more dominant than the \ac{NLoS} ones, which can be contributed by the signals transmitted by the sidelobes of the directional antenna at the \ac{BS}. We can also observe that the \ac{RMS} delay spread values are slightly lower for the Environment~I with respect to the Environment~II since the the channel exhibits more clearance for the first case.  \reviewXose{Finally, a noticeable increase of both the mean and the variance of the \ac{RMS} delay spread can be appreciated in the Environment~II when the \ac{UAV} crosses the media building for the lowest flight (i.e., the $15$\,m height one), due to the change from \ac{LoS} propagation conditions to \ac{OLoS}.}

\begin{table}
	\centering
	\begin{small}
		\resizebox{\columnwidth}{!}{
		\begin{tabular} {ccccc}
			\toprule
			Height & Env.~I & Env.~I & Env.~II & Env.~II\\
			{}[m] & Omnidir. ant. & Dir. ant & Omnidir. ant. & Dir. ant\\
			\midrule
                        \reviewXose{$15$ (LoS)}&$\left(-6.81,0.16\right)$&$\left(-6.56,0.26\right)$&\reviewXose{$\left(-6.92,0.12\right)$}&\reviewXose{$\left(-6.75,0.19\right)$}\\
\reviewXose{$15$ (OLoS)}&\reviewXose{--}&\reviewXose{--}&\reviewXose{$\left(-6.66,0.27\right)$}&\reviewXose{$\left(-6.52,0.32\right)$}\\
$25$&$\left(-6.81,0.16\right)$&$\left(-6.79,0.37\right)$&$\left(-6.63,0.16\right)$&$\left(-6.86,0.25\right)$\\
$35$&$\left(-6.68,0.25\right)$&$\left(-6.66,0.28\right)$&$\left(-6.63,0.16\right)$&$\left(-6.86,0.25\right)$\\
$45$&$\left(-6.68,0.25\right)$&$\left(-6.73,0.21\right)$&$\left(-6.63,0.16\right)$&$\left(-6.86,0.25\right)$\\
$60$&$\left(-6.68,0.25\right)$&$\left(-6.80,0.20\right)$&$\left(-6.63,0.16\right)$&$\left(-6.86,0.25\right)$\\
$75$&$\left(-6.81,0.16\right)$&$\left(-6.73,0.21\right)$&$\left(-6.63,0.16\right)$&$\left(-6.43,0.26\right)$\\
$90$&$\left(-6.81,0.16\right)$&$\left(-6.73,0.21\right)$&$\left(-6.63,0.16\right)$&$\left(-6.43,0.26\right)$\\
$105$&$\left(-6.81,0.16\right)$&$\left(-6.73,0.21\right)$&$\left(-6.63,0.16\right)$&$\left(-6.43,0.26\right)$\\
			\bottomrule
		\end{tabular}
	}
	\end{small}
	\caption{Parameters of the normal distributions used to fit the \ac{RMS} delay spread \minorReviewXose{(in $\log_{10}$\,[s])} for the different measurement environments and transmit antennas. The parameters are specified in the format $(\mu,\sigma^2)$, where $\mu$ denotes the mean and $\sigma^2$ the variance.\label{table:DelaySpread}}
\end{table}

\subsection{Root Mean Square Doppler Frequency Spread}\label{sec:rmsDopplerFrequencySpread}

The Doppler frequency spread is inversely related with the channel coherence time and hence could affect both the maximum usable frame size as well as the duplexing method. It also leads to \ac{ICI}, hence making it necessary to include \ac{ICI} cancellation methods at the receiver or more advanced channel equalization techniques. The \ac{RMS} Doppler Frequency Spread is calculated as the second-order central moment of the Doppler \ac{PSD}, in a analogous way as that described by \cref{eq:normalized-pdp,eq:n-moment-pdp,eq:rms-delay-spread}. Let us define the normalized Doppler \ac{PSD} for the $i$th frame, by applying a similar strategy as for the delay, as
\begin{equation}
\label{eq:normalized-doppler}
\tilde{D}_i(\nu) = \frac{D_i(\nu)}{\int_{-\infty}^{\infty}D_i(\nu) \dif \nu} =
\frac{\sum_{l=1}^{M}  \left|\alpha_{i,l}\right|^2 \delta\left(\nu-\nu_{i,l}\right)}
{\sum_{l=1}^{M} \left|\alpha_{i,l}\right|^2},
\end{equation}
where $D_i(\nu)$ is defined as in \cref{eq:doppler-psd-sage}. From the result in \cref{eq:normalized-doppler}, the $n$th  moment of the Doppler is
\begin{equation}
\label{eq:n-moment-doppler}
\mathbb{E}_i[\nu^n] = \int_{-\infty}^{\infty}   \tilde{D}_i(\nu) \nu^n \dif \nu =
\frac{\sum_{l=1}^{M}  \left|\alpha_{i,l}\right|^2 \nu_{i,l}^n}
{\sum_{l=1}^{M} \left|\alpha_{i,l}\right|^2},
\end{equation}
and the \ac{RMS} Doppler frequency spread is defined as
\begin{equation}
\label{eq:rms-doppler-spread}
R_i = \sqrt{\mathbb{E}_i[\nu^2] - \mathbb{E}_i[\nu]^2}.
\end{equation}

\begin{figure}[t!]
	\centering
	\begin{subfigure}[t]{\linewidth}
		\centering
		\includegraphics[width=\columnwidth]{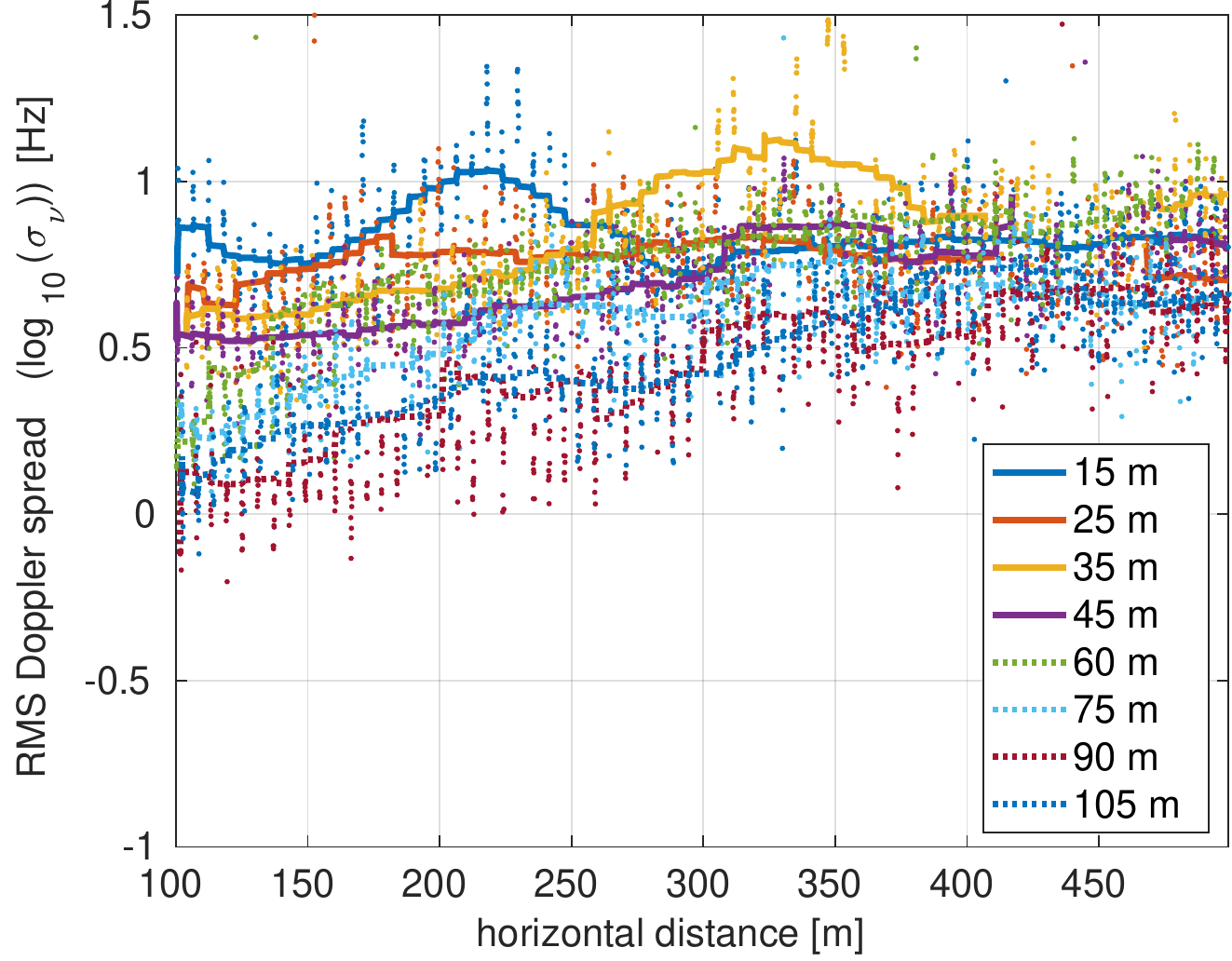}
		\caption{RMS Doppler frequency spread.}
		\label{Fig:dops-2500MHzLoS-omni}
	\end{subfigure}
	\par\medskip
	\begin{subfigure}[t]{\linewidth}
		\centering
		\includegraphics[width=\columnwidth]{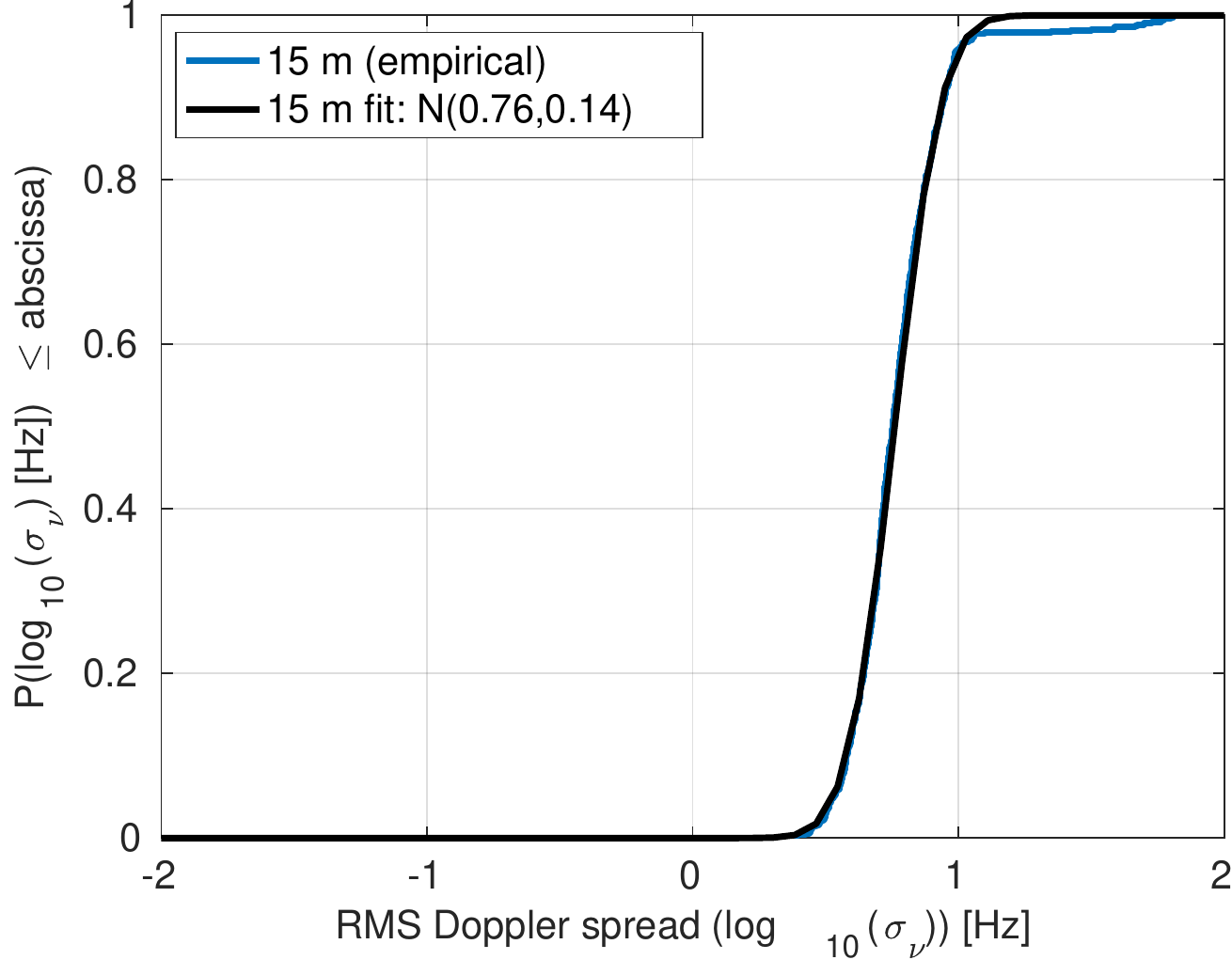}
		\caption{RMS Doppler frequency spread fitting for the flight height $15$\,m.}
		\label{Fig:cfdops-2500MHzLoS-omni}
	\end{subfigure}
	\caption{RMS Doppler frequency spread for different heights and sample fittings when the omnidirectional antenna is used at the transmitter and the Environment~I is considered.}
	\label{fig:figureDFS}
\end{figure}

\cref{Fig:dops-2500MHzLoS-omni} shows the RMS Doppler frequency spread values for the sample case. It can be observed that the \ac{RMS} Doppler Frequency Spread values tend to decrease as the flight height increases, specially when the \ac{UAV} is close to the \ac{BS}.
\cref{Fig:cfdops-2500MHzLoS-omni} shows a sample proposed fitting for one of the obtained empirical \acp{CDF}, being the obtained results for all the cases summarized in \cref{table:DopplerSpread}. In all cases, normal distributions are used for the fittings. The obtained results show that the \ac{RMS} Doppler Frequency spread decreases when the flight height increases, and that in general it is lower for the Environment~I when the omnidirectional antenna is used, being the results more similar for the \ac{BS} antenna case. Furthermore, the use of the \ac{BS} antenna can slightly increase the \ac{RMS} Doppler frequency spread. \reviewXose{\cref{table:DopplerSpread} also reveals an increase of the mean and the variance of the \ac{RMS} Doppler frequency spread when the \ac{UAV} changes from \ac{LoS} propagation conditions to \ac{OLoS} after crossing the media building when the flight height is $15$\,m.}

\begin{table}
	\centering
	\begin{small}
		\resizebox{\columnwidth}{!}{
		\begin{tabular} {ccccc}
			\toprule
			Height & Env.~I & Env.~I & Env.~II & Env.~II\\
			{}[m] & Omnidir. ant. & Dir. ant & Omnidir. ant. & Dir. ant\\
			\midrule
                        \reviewXose{$15$ (LoS)}&$\left(0.76,0.14\right)$&$\left(1.40,0.16\right)$&\reviewXose{$\left(0.90,0.15\right)$}&\reviewXose{$\left(1.12,0.11\right)$}\\
\reviewXose{$15$ (OLoS)}&\reviewXose{--}&\reviewXose{--}&\reviewXose{$\left(1.08,0.23\right)$}&\reviewXose{$\left(1.23,0.29\right)$}\\
$25$&$\left(0.76,0.14\right)$&$\left(0.99,0.21\right)$&$\left(1.06,0.30\right)$&$\left(0.98,0.29\right)$\\
$35$&$\left(0.86,0.20\right)$&$\left(1.18,0.20\right)$&$\left(1.06,0.30\right)$&$\left(1.08,0.29\right)$\\
$45$&$\left(0.76,0.14\right)$&$\left(1.18,0.20\right)$&$\left(1.06,0.30\right)$&$\left(0.98,0.29\right)$\\
$60$&$\left(0.79,0.21\right)$&$\left(0.99,0.21\right)$&$\left(1.03,0.30\right)$&$\left(0.98,0.29\right)$\\
$75$&$\left(0.50,0.22\right)$&$\left(0.99,0.21\right)$&$\left(0.87,0.23\right)$&$\left(0.80,0.27\right)$\\
$90$&$\left(0.50,0.22\right)$&$\left(0.99,0.21\right)$&$\left(0.50,0.15\right)$&$\left(0.98,0.29\right)$\\
$105$&$\left(0.50,0.22\right)$&$\left(0.99,0.21\right)$&$\left(0.50,0.15\right)$&$\left(0.80,0.27\right)$\\
			\bottomrule
		\end{tabular}
	}
	\end{small}
	\caption{Parameters of the normal distributions used to fit the \ac{RMS} Doppler frequency spread for the different measurement environments and transmit antennas. The parameters are specified in the format $(\mu,\sigma^2)$, where $\mu$ denotes the mean and $\sigma^2$ the variance.\label{table:DopplerSpread}}
\end{table}

\subsection{Ricean K-factor}\label{sec:riceanKFactor}

In principle, for both of the considered measurement environments, mainly (possibly obstructed) \ac{LoS} propagation conditions are assumed during most of the trajectory. Hence, one of the paths exhibits a much higher power level than the others \reviewXose{most of the time}. In this situation\minorReviewXoseRemoveText{s}, the fluctuations of that path gain can be assumed to follow a Ricean distribution, which is characterized by a single parameter, namely the Ricean K-factor \cite{ArtigoMedidasTren_WCMC2017}. More specifically, the K-factor is
the ratio of the power in the \ac{LoS} component or dominant
component to the power in the \ac{NLoS} or the other multipath
components \cite{7489014}. The classical moment-based method proposed in \cite{greensteinmethod} was used to calculate the K-factor.

\reviewXoseRemoveText{Fig.~9a shows the K-factor values for the sample case. It can be seen that, in general, the K-factor decreases with the distance between the UAV and the BS due to the power decrease of the LoS component, except for the lowest flights. In those cases, the K-factor increases slightly with the distance or remains approximately constant. This is due to the presence of rich MPCs close to the BS. Due to this, for the areas close to the BS, the maximum K-factor values are obtained for the highest flights, whereas for the largest distances the low-height flights exhibit higher K-factors. Fig.~9b shows an example of proposed fitting for one of the obtained empirical CDF, being all the obtained results summarized in Table~VI. In all cases, Normal distributions are used for the fittings. From the results, it can be seen that, when the omnidirectional antenna is used, the middle-height flights exhibit slightly lower results, whereas the extreme (upper and lower) flights increase the K-factor, which is coherent with the effects shown in Fig.~9a. When the directional antenna is used at the BS, the K-factor results are slightly decreased, specially for the lowest and highest flights. However, for the mid-height flights in the Environment~II, the K-factor is increased by the use of the directional antenna, probably due to the effect of the radiation pattern. One special case to mention is the $15$\,m height flight for the Environment~II, regardless of the transmitter antenna. In this case, the K-factor is noticeably lower than for other flights. The reason is that a drop in the K-factor value occurs when the horizontal distance is about $250$\,m. This is coherent with the received power drop for this specific flight mentioned in Section~IV-A and supports the explanation that it is due to the partial loss of the LoS propagation conditions.}

\begin{figure}[t!]
	\centering
	\begin{subfigure}[t]{\linewidth}
		\centering
		\includegraphics[width=\columnwidth]{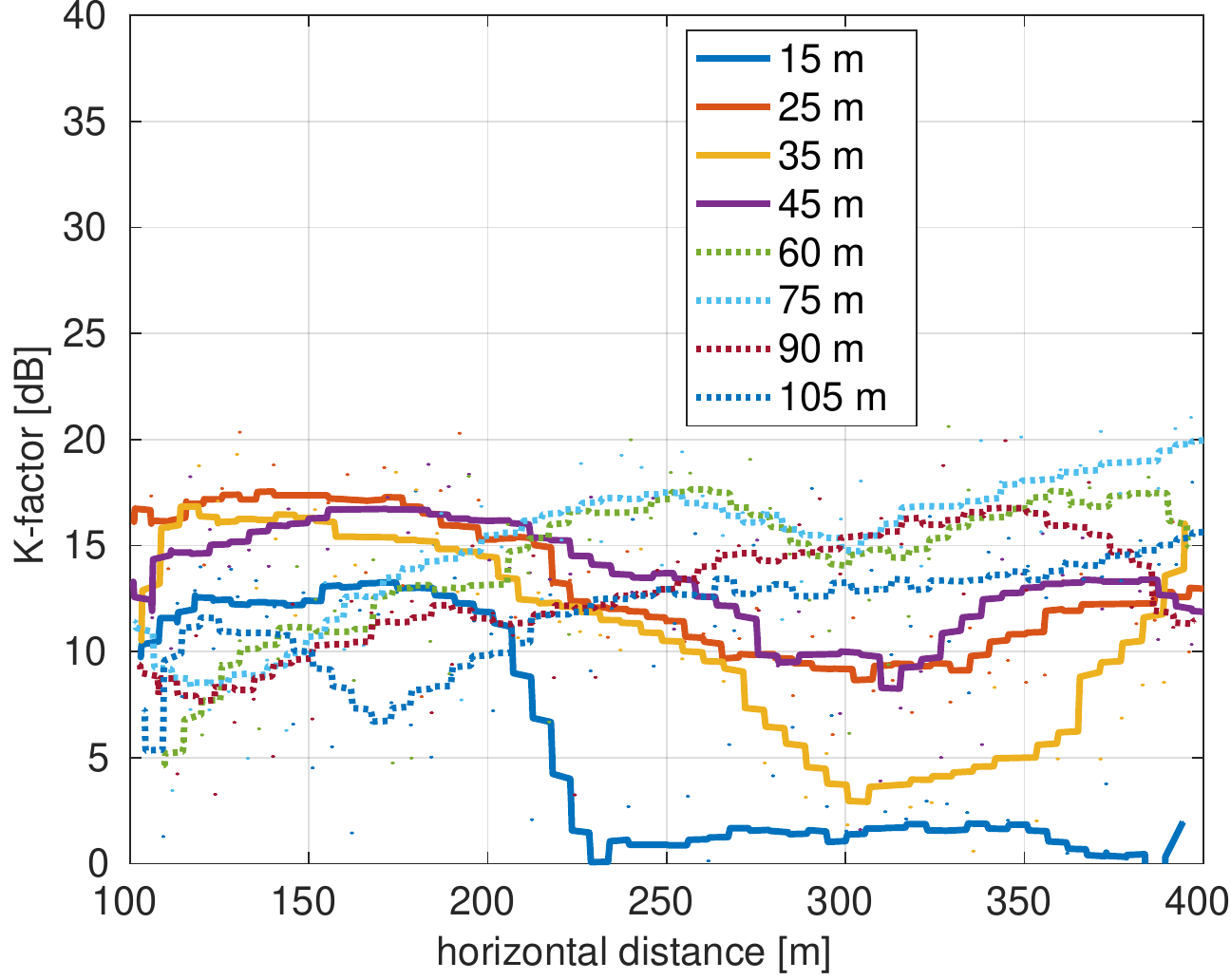}
		\caption{\reviewXose{K-factor}. \reviewCommentXose{This figure was changed.}}
		\label{Fig:kf-2500MHzNLoS-bs}
	\end{subfigure}
	\par\medskip
	\begin{subfigure}[t]{\linewidth}
		\centering
		\includegraphics[width=\columnwidth]{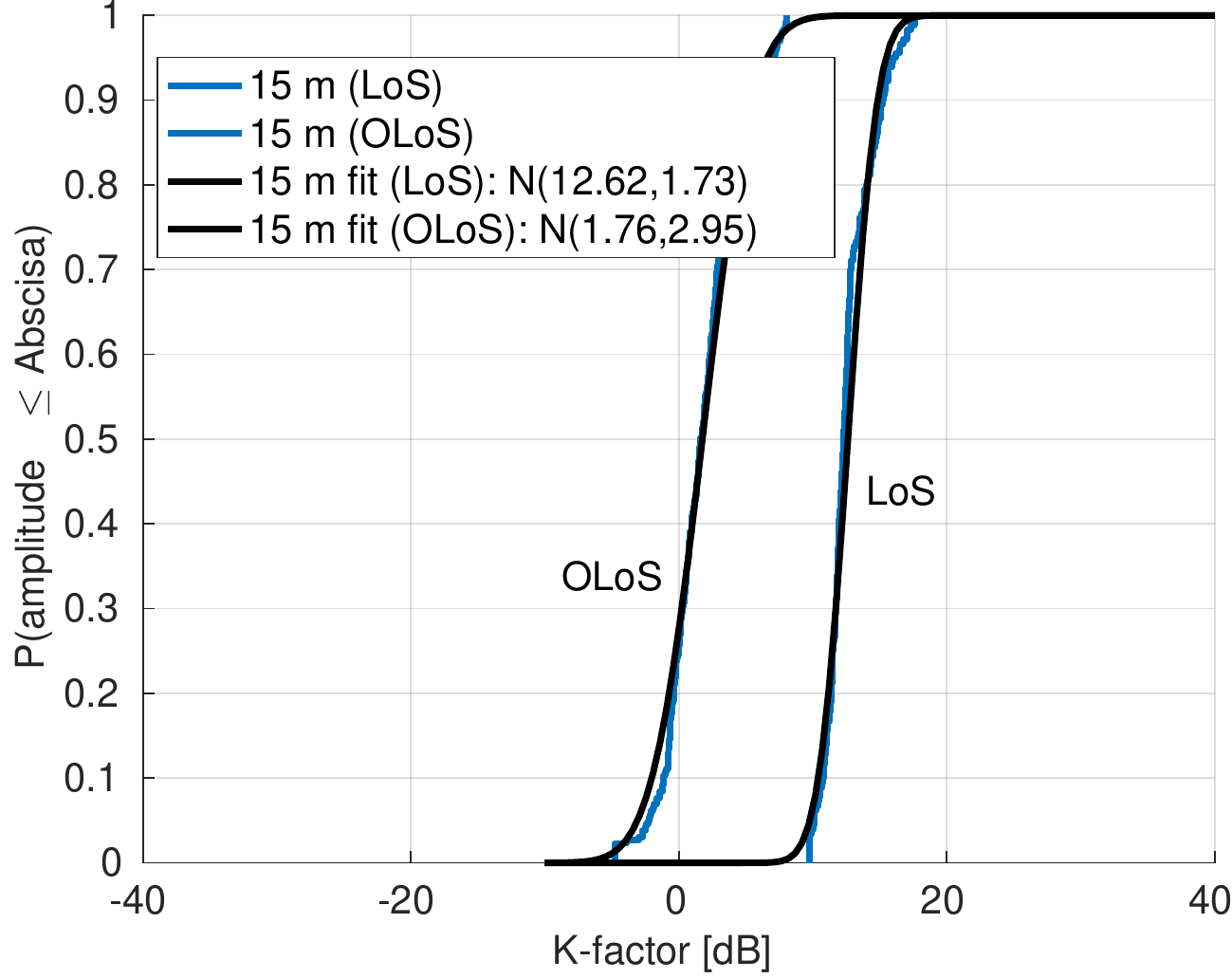}
		\caption{K-factor fitting\reviewXose{s} for the flight height $15$\,m. \reviewCommentXose{This figure was changed.}}
		\label{Fig:fckf-2500MHzNLoS-bs}
	\end{subfigure}
	\caption{K-factor values for different heights and sample fittings when the \reviewXose{\ac{BS}} antenna is  used at the transmitter and the \reviewXose{Environment~II} is considered.}
	\label{fig:figureKFactor}
\end{figure}

\reviewXose{As in previous sections, the K-factor calculation is also divided into two parts when the flight height is $15$\,m for the Environment~II, labeled as \ac{LoS} and \ac{OLoS}, respectively. In fact, in order to illustrate this effect, \cref{Fig:kf-2500MHzNLoS-bs} shows the K-factor at different flight heights for the Environment~II considering the \ac{BS} antenna at the transmitter. Noticeably, different behaviors can be observed with the flight height. Firstly, there is a sharp drop in the $15$\,m case when the \ac{UAV} crosses the media building, hence moving from the \ac{LoS} area to the \ac{OLoS} one. Secondly, it can be seen that for moderate flight heights (below $45$\,m), the general trend of the K-Factor is to decrease with the horizontal distance. The main reason for this behavior is that the density of buildings close to the last part of the flight is increased, which results in more reflections close in power to the \ac{LoS} component approaching the \ac{UAV}. Indeed, this effect is also appreciated for the case of the Environment~I. Furthermore, for moderate flight heights (below $45$\,m--$60$\,m) in Environment~II, there is a local minimum of the K-Factor when the horizontal distance is around $325$\,m, as shown in \cref{Fig:kf-2500MHzNLoS-bs}, which corresponds to the area more close to the library building. This way, in these areas, the reflections caused by the library building became more similar in power to the \ac{LoS} component (this effect can also be appreciated in \cref{Fig:pdp-2500MHzNLoS-omni-15}) and decrease the K-Factor. Finally, for higher flights, the K-Factor increases with the distance in general when the \ac{BS} antenna is used, as shown in \cref{Fig:kf-2500MHzNLoS-bs}. This is due to the effect of the radiation pattern of the \ac{BS} antenna. As explained in \cref{sec:PathLoss}, the \ac{UAV} is expected to fall out from the main radiation lobe for low values of the horizontal distance, whereas larger distances may allow to receive some contributions from the main radiation lobe, which results in an increased K-Factor. This is not the case when the omnidirectional antenna is used. In this case, the general trend of the K-Factor is to be decreased with the horizontal distance increasing regardless of the flight height. {\cref{Fig:fckf-2500MHzNLoS-bs} shows the sample fittings for the obtained empirical \acp{CDF} when the flight height is $15$\,m (in both \ac{LoS} and \ac{OLoS} cases).}
		
\reviewXose{All the obtained K-factor results are summarized in \cref{table:KF}}.  In all cases, Normal distributions are used for the fittings. From the results, it can be seen that, when the omnidirectional antenna is used, the middle-height flights exhibit slightly lower results, whereas the highest flights increase the K-factor, which may be partially caused due to the effect of the library building (consequently with the effects shown in \cref{Fig:kf-2500MHzNLoS-bs}). Since this building is around $63$\,m height, it is likely to affect less to the highest flights. When the directional antenna is used at the \ac{BS}, the K-factor results are in general decreased, specially for the highest flights. As in the example shown in \cref{Fig:kf-2500MHzNLoS-bs}, the K-Factor for the \ac{BS} case and high flights increases in general with the horizontal distance, but some values for low horizontal distances can be very reduced. This leads to a lower mean value and an increased variance for the fits in general. Finally, it can be seen that the K-factor does not exhibit as much dependency with the flight height as it could be expected. Indeed, although the channel is expected to be more \ac{LoS}-alike for higher flights, due to the existence of high buildings in the measurement environments (e.g., the library building or the buildings close to the entrance to the university campus), strong reflections can still be observed at high flight heights, which causes the variation of the K-factor to be not so significant. Even more, in some cases those reflections in high buildings can only be permanently appreciated for high flights, since in lower flights the reflections can be temporarily blocked by other elements of the ground environment in the vicinity of the \ac{UAV}.}

\begin{table}
\centering
\begin{small}
	\resizebox{\columnwidth}{!}{
	\begin{tabular} {ccccc}
		\toprule
		Height & Env.~I & Env.~I & Env.~II & Env.~II\\
		{}[m] & Omnidir. ant. & Dir. ant & Omnidir. ant. & Dir. ant\\
		\midrule
                \reviewXose{$15$ (LoS)}&$(15.82,2.25)$&$(5.54,3.13)$&\reviewXose{$(14.58,1.78)$}&\reviewXose{$(12.62,1.73)$}\\
\reviewXose{$15$ (OLoS)}&\reviewXose{--}& \reviewXose{--}& \reviewXose{$(4.58,2.41)$}&\reviewXose{$(1.76,2.95)$}\\
$25$&$(14.46,1.96)$&$(11.37,3.69)$&$(10.92,2.51)$&$(12.00,4.12)$\\
$35$&$(12.67,3.40)$&$(9.30,3.96)$&$(12.28,2.37)$&$(9.96,5.56)$\\
$45$&$(13.50,2.53)$&$(8.19,3.29)$&$(11.54,3.01)$&$(13.29,2.96)$\\
$60$&$(11.64,3.85)$&$(11.23,4.59)$&$(10.89,3.70)$&$(12.91,5.30)$\\
$75$&$(14.35,2.94)$&$(10.30,3.36)$&$(12.72,3.40)$&$(15.45,3.28)$\\
$90$&$(15.01,4.82)$&$(10.00,3.29)$&$(16.28,3.61)$&$(11.76,4.69)$\\
$105$&$(14.02,4.98)$&$(9.19,4.49)$&$(15.62,2.37)$&$(11.13,3.95)$\\
		\bottomrule
	\end{tabular}
	}
\end{small}
\caption{Parameters of the normal distributions used to fit the K-factor for the different measurement environments and transmit antennas. The parameters are specified in the format $(\mu,\sigma^2)$, where $\mu$ denotes the mean and $\sigma^2$ the variance.\label{table:KF}}
\end{table}

\section{Conclusions}\label{sec:Conclusions}

In this paper, the air-to-ground channel for low-height small-sized \acp{UAV} was studied based on a systematic measurement campaign including horizontal flights at different heights and two different suburban environments. Two antennas were used at the transmitter, which are \textit{(a)} an omnidirectional antenna, which allows for obtaining an accurate characterization of the channel characteristics and \textit{(b)} a typical \ac{BS} directional antenna, which gives insights on the behavior for \reviewXoseRemoveText{a more realistic deployment}\reviewXose{an individual sector of a \ac{BS}}. \secondMinorReviewXoseRemoveText{Firstly, t}\secondMinorReviewXose{T}he channel for the different measurement environments was characterized based on the path loss, shadow fading, \ac{PDP}, Doppler frequency \ac{PSD}, \ac{RMS} delay spread, \ac{RMS} Doppler frequency spread and the K-factor. From the results, it can be seen that the ground scenario can severely affect the path loss exponent, specially for low heights. The choose of the antenna type as well as its orientation (in the case of directive ones) is also very relevant depending on the flight height range to be considered. The shadow fading does not show a high dependency with the distance between the \ac{UAV} and the \ac{BS}, but it is reduced with the flight height \minorReviewXose{increasing} when the omnidirectional antenna is used. Due to the effect of the sidelobes, this effect cannot be observed when the directional antenna is considered. \ac{RMS} delay and Doppler frequency spreads do not show a strong dependency with the flight height, although they tend to decrease slightly for the highest flight values considered. The use of the directional antenna can help to slightly reduce the \ac{RMS} delay spread for mid-height flights, leading to decreased error probability due to \minorReviewXose{less} delay dispersion. The ground elements can still affect the \ac{RMS} delay spread even for high flights. \reviewXose{Finally, the K-factor, as well as the path loss, are severely affected by the ground elements and the radiation pattern of the antenna at the \ac{BS}.} \reviewXoseRemoveText{ is maximum for high flights when the distance to the BS is low, although it tends to decrease with the distance to the BS}\reviewXose{Due to the reflections in high buildings are still noticeable when the \ac{UAV} flight height is high, the effect of the flight height on the K-factor could be not so obvious in this kind of environments.}

Summing up the propagation results, we can see that the channel characteristics are \secondMinorReviewXoseRemoveText{quite}affected by the ground environment, even for not so low flight heights. Furthermore, it can be seen that the results are influenced by the antenna used at the \ac{BS}. It can be seen that the use of a directional \ac{BS} antenna, as usual for ground developments (even for $0^{\circ}$ tilt), limits  the \reviewXose{flight} height \reviewXose{and distance} range\reviewXose{s}, and can influence the time and frequency coherence of the channel up to some extent. \secondReviewXose{This way, the use of terrestrial \ac{LTE}/\ac{5G} deployments to serve \acp{UAV} will probably be subjected to the careful consideration of the antennas configuration at the \ac{BS}. In this sense, for a typical deployment based on the use of several directional antennas, it may be required to add more antennas to cover a larger height range. In this situation, the provided statistical channel model for the omnidirectional antenna at the \ac{BS} would be a good approximation to the channel model between the \ac{UAV} and a \ac{BS} with several directional antennas for the range of height covered by those directional antennas. Hence, the provided channel model can be used to estimate the actual performance of \ac{LTE}/\ac{5G} communication systems for serving \acp{UAV}, as we have already done in \cite{ArtigoSimulacionsParaUAV_EuCAP2020}.}
	
\secondReviewXose{Note that, even though we have shown that the analyzed channel characteristics vary with the height and the distance between the \ac{UAV} and the \ac{BS}, for most of them the variation is not so significant and the parameters of most of the provided statistical distributions change indeed within a reduced range for realistic flight heights according to the regulations. The channel characteristics that are more dependent with the flight height and the \ac{UAV}-\ac{BS} distance are the path loss and the Ricean K-Factor. Except for the cases where this limitation comes from the radiation pattern of the antenna (which should not be the case for a proper \ac{BS} deployment), indeed the variation is not so directly related with the flight height, but with the distribution of the elements of the ground environment. This way, depending on the distribution and height of the buildings and other architectural elements in the environment, it could be not straightforward to estimate a parameter such as the K-Factor. For example, the K-Factor can be decreased when the flight height increases (which seems not intuitive in principle) since for higher flights the reflections from tall buildings are not blocked by short buildings close to the \ac{UAV} and hence they become more stable. This  leads to the conclusion that time-variant channel models are required for this kind of scenarios, which can predict how the distribution of the \acp{MPC} change with time or, equivalently, with the position of the \ac{UAV} with respect to the ground environment, which is our current work line for the next future works.}



%



\bibliographystyle{IEEEtran}
\bibliography{IEEEabrv,ArtigoCaracteristicasCanleComunicacionsUAV_2019-CombinedTAP}



\if\printBiographies1


\begin{IEEEbiography}[
	{\includegraphics[width=1in,height=1.25in,clip,keepaspectratio]{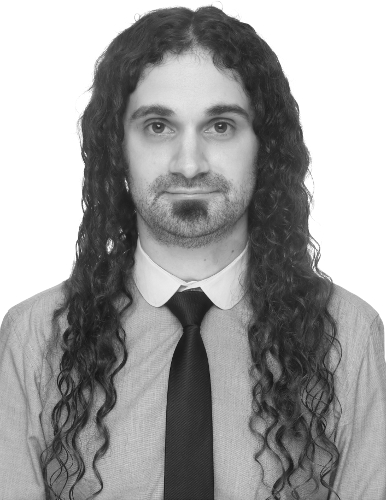}}
	]{José Rodríguez-Piñeiro} received the B.Sc. on  Telecommunications and the M.Sc. Degree in Signal Processing Applications for Communications from the University of Vigo (Pontevedra, Spain), in 2009 and 2011, respectively. Between June 2008 and July 2011, he was a researcher at the Department of Signal and Communications, University of Vigo (Pontevedra, Spain). From October 2011 he was a researcher at the Group of Electronics Technology and Communications of the University of A Coruña (UDC), obtaining his Ph.D. degree with the distinction “Doctor with European Mention” in 2016. After obtaining his Ph.D. degree with the he continued working as a Postdoctoral researcher at the same group until July 2017. On August 2017 he joined the College of Electronics and Information Engineering, Tongji University (P.R. China), becoming an Assistant Professor in 2020. From November 2012 he also collaborates with the Department of Power and Control Systems, National University of Asunción (Paraguay) in both teaching and research. He is the coauthor of more than 50 papers in peer-reviewed international journals and conferences. He is also a member of the research team in more than 25 research projects funded by public organizations and private companies. He was awarded with 6 predoctoral, postdoctoral and research stay grants. His research interests include experimental evaluation of digital mobile communications, especially for high mobility environments, including terrestrial and aerial vehicular scenarios.
\end{IEEEbiography}



\begin{IEEEbiography}[{\includegraphics[width=1in,height=1.25in,clip,keepaspectratio]{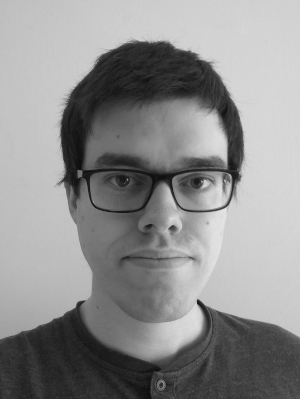}}
	]{Tomás Domínguez-Bolaño} received the B.S degree in Computer Engineering and the Ph.D. in Computer Engineering (with the distinction ``Doctor with European Mention'') from the University of A Coruña, A Coruña, Spain, in 2014 and 2018, respectively. Since 2014 he has been with the Group of Electronics Technology and Communications. In 2018 he was a Visiting Scholar with Tongji University, Shanghai, China. He is an author of more than 15 papers in peer-reviewed international journals and conferences. He was awarded with a predoctoral grant and two research-stay grants. His research interests include channel measurements, parameter estimation and modeling and experimental evaluation of wireless communication systems.
\end{IEEEbiography}



\begin{IEEEbiography}[
	{\includegraphics[width=1in,height=1.25in,clip,keepaspectratio]{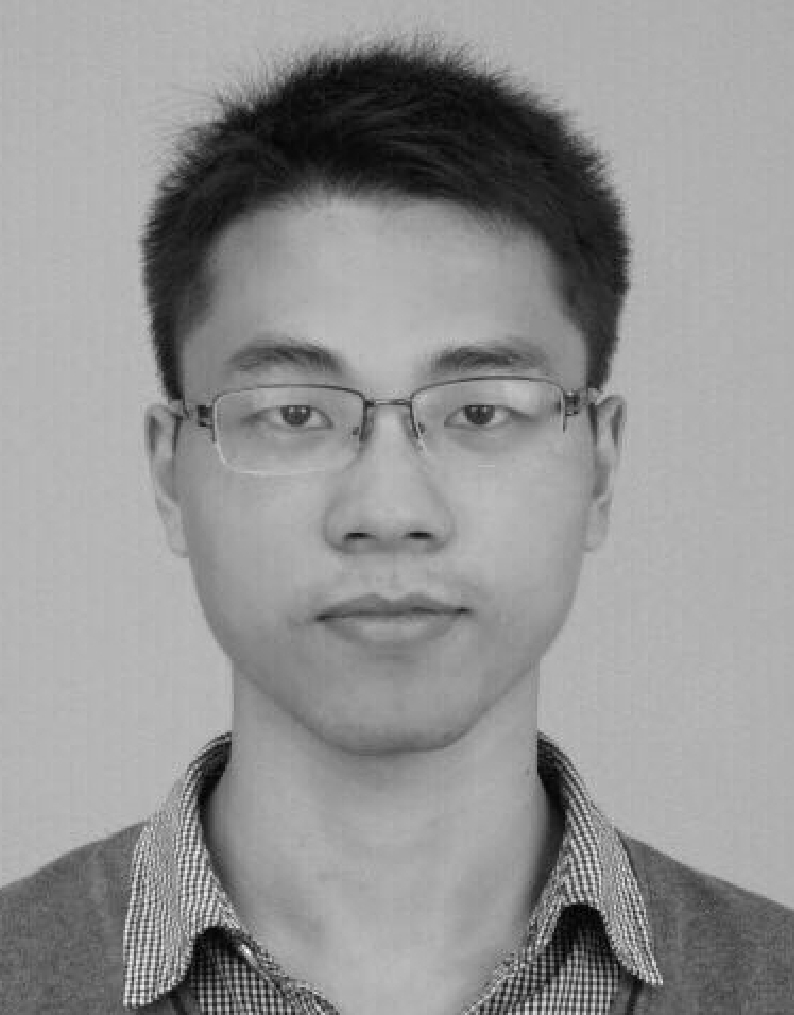}}
	]{Xuesong Cai} received the B.S. degree and the Ph.D. degree (Hons.) from Tongji University, Shanghai, China, in 2013 and 2018, respectively. In 2015, he conducted a three-month internship with Huawei Technologies, Shanghai, China. He was also a Visiting Scholar with Universidad Polit\'ecnica de Madrid, Madrid, Spain in 2016. From 2018-2020, he was a postdoctoral research fellow with the APMS section, Department of Electronic Systems, Aalborg University (AAU), Aalborg, Denmark. Since April 2020, he has been a postdoctoral fellow with the Wireless Communication Networks Section, Department of Electronic Systems, AAU, cooperating with Nokia Bell Labs. His research interests include propagation channel measurement, high-resolution parameter estimation, channel characterization, channel modeling and over-the-air emulation for wireless communications.
	
	Dr. Cai was a recipient of the Chinese National Scholarship for Ph.D. Candidates and Excellent Student award in 2016, the Excellent Student award and the ``ZTE Fantastic Algorithm'' award in 2017, the Outstanding Doctorate Graduate awarded by Shanghai Municipal Education Commission and ``ZTE Blue Sword-Future Leaders Plan'' in 2018, and the ``Seal of Excellence'' with the European Horizon 2020’s Marie Skłodowska-Curie actions call in 2019. 


\end{IEEEbiography}



\begin{IEEEbiography}[
	{\includegraphics[width=1in,height=1.25in,clip,keepaspectratio]{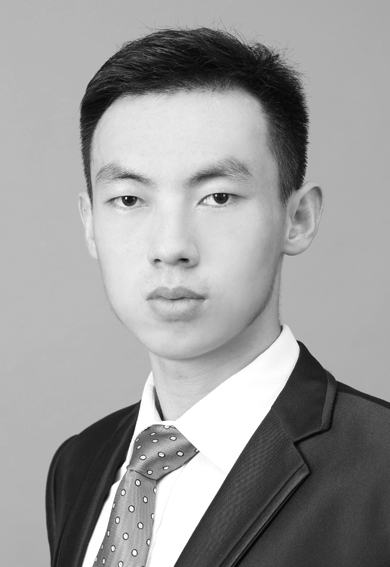}}
	]{Zeyu Huang} received the B.S. degree in electronics and information engineering from Tongji University, Shanghai, China, in 2015. Since 2017 he was with the College of Electronics and Information Engineering, Tongji University (P.R. China), obtaining his M.Sc. degree in 2020. He was accepted as a Ph.D. student at Technische Universit{\"a}t Wien (Vienna, Austria) in 2020. He co-authored 3 papers in peer-reviewed international journals and conferences and he was a member of the research team in 7 research projects. His research interests include communications channel measurement and simulation, parameter estimation of communication channels, channel modeling and the use of meta-materials for communications.
\end{IEEEbiography}



\begin{IEEEbiography}[
	{\includegraphics[width=1in,height=1.25in,clip,keepaspectratio]{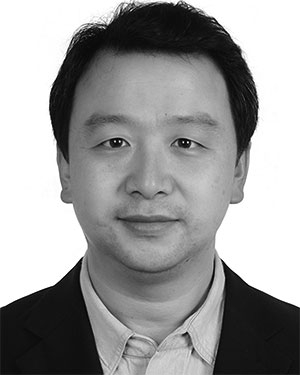}}
	]{Xuefeng Yin} (S’01–M’06) received the Bachelor's degree in optoelectronics engineering from Huazhong University of Science and Technology, Wuhan, China, in 1995, and the Master of Science degree in digital communications and the Ph.D. in wireless communications from Aalborg University, Aalborg, Denmark, in 2002 and 2006, respectively. From 2006 to 2008, he was an Assistant Professor with Aalborg University. In 2008, he joined the College of Electronics and Information Engineering, Tongji University, Shanghai, China. He became a Full Professor in 2016 and has been the Vice Dean of the college since then. His research interests include high-resolution parameter estimation for propagation channels, measurement-based channel characterization and stochastic modeling for 5G wireless communications, channel simulation based on random graph models, radar signal processing, and target recognition. He has authored or coauthored more than 150 technical papers and coauthored the book Propagation Channel Characterization, Parameter Estimation and Modeling for Wireless Communications (Wiley, 2016).
\end{IEEEbiography}


\else

\fi


\end{document}